\documentclass[12pt]{article}

\usepackage{epsfig}

\newcommand{\sect}[1]{\setcounter{equation}{0}\section{#1}}
\renewcommand{\theequation}{\arabic{section}.\arabic{equation}}
\renewcommand{\Re}{{\rm Re}}
\renewcommand{\Im}{{\rm Im}}

\def\be{\begin{equation}}
\def\ee{\end{equation}}
\def\ba{\begin{eqnarray}}
\def\ea{\end{eqnarray}}

\begin{document}

\thispagestyle{empty}

\hfill{}

\hfill{}

\hfill{CERN-TH/2001-293}


\hfill{hep-th/0110258}

\vspace{32pt}

\begin{center} 
\textbf{\Large Generalized Weyl Solutions} 

\vspace{40pt}

Roberto Emparan$^{a,}$\footnote{Also at Departamento de F{\'\i}sica
Te\'orica, Universidad del Pa{\'\i}s Vasco, E-48080, Bilbao, Spain.}
and
Harvey S.~Reall$^b$

\vspace{12pt}

$^a$\textit{Theory Division, CERN}\\
\textit{CH-1211 Geneva 23, Switzerland}\\
\vspace{6pt}
$^b$\textit{Physics Department, Queen Mary College\\
Mile End Road, London E1 4NS, United Kingdom}\\
\end{center}

\vspace{40pt}

\begin{abstract}

It was shown by Weyl that the general static axisymmetric solution of
the vacuum Einstein equations in four dimensions is given in terms of a
single axisymmetric solution of the Laplace equation in
three-dimensional flat space. Weyl's construction is generalized here to
arbitrary dimension $D \ge 4$. The general solution of the
$D$-dimensional vacuum Einstein equations that admits $D-2$ orthogonal
commuting non-null Killing vector fields is given either in terms of
$D-3$ independent axisymmetric solutions of Laplace's equation in
three-dimensional flat space or by $D-4$ independent solutions of
Laplace's equation in two-dimensional flat space. Explicit examples of
new solutions are given. These include a five-dimensional asymptotically
flat ``black ring'' with an event horizon of topology $S^1 \times S^2$
held in equilibrium by a conical singularity in the form of a disc. 

\end{abstract}

\newpage

\sect{Introduction}

Exact solutions play an important role in General Relativity. 
Examining properties of exact solutions has led to deep insights into
the nature of spacetime that would have been hard to arrive at by
other means. For example, much of the progress made in understanding
properties of black holes in the 60's and 70's relied on the existence
of the Kerr-Newman solution. The Standard Model of cosmology is built
on Friedman-Robertson-Walker solutions. Examining properties of
Bianchi cosmologies has led to insight into how inflation dissipates 
anisotropy.

Much effort has been devoted to developing techniques for finding exact
solutions in four dimensions \cite{kramer:80, wald:84}. One of the
earliest results in this direction was obtained by Weyl \cite{weyl:17},
who found the general static axisymmetric solution of the vacuum
Einstein equations:
\be
\label{eqn:4dweylmetric}
 ds^2 = -e^{2U} dt^2 + e^{-2U} \left( e^{2 \gamma}(dr^2 + dz^2) + r^2
 d\phi^2 \right),
\ee
where $U(r,z)$ is an arbitrary axisymmetric solution of Laplace's
equation in a three-dimensional {\it flat} space with metric
\be
 ds^2 = dr^2 + r^2 d\phi^2 + dz^2,
\ee
and $\gamma$ satisfies
\be
 \frac{\partial \gamma}{\partial r} = r \left[ \left(\frac{\partial
 U}{\partial r} \right)^2 - \left(\frac{\partial
 U}{\partial z} \right)^2 \right],
\ee
\be
 \frac{\partial \gamma}{\partial z} = 2 r \frac{\partial U}{\partial
 r} \frac{\partial U}{\partial z}.
\ee
The solution of these equations is given by a line integral. Since $U$
is harmonic, it can be regarded as a Newtonian potential produced by
certain (axisymmetric) sources. For example, the Schwarzschild
solution corresponds to taking the source for $U$ to be a thin rod on
the $z$-axis with mass $1/2$ per unit length.  

Nowadays, interest in solutions of General Relativity is no longer
restricted to four dimensions. Many interesting solutions of higher
dimensional supergravity theories have been found. In spite of this,
there are basic questions concerning the nature of gravity in higher
dimensions that remain unanswered. In four dimensions, it can be proved
that each connected component of the event horizon of an asymptotically
flat spacetime satisfying the dominant energy condition has topology
$S^2$ \cite{hawking:73}. The proof relies on the Gauss-Bonnet theorem
applied to a constant time slice through the horizon, and is therefore
invalid in higher dimensions. A different approach to rule out
non-spherical topologies is based on the notion of `topological
censorship' \cite{gannon:75}. However, this argument is typically
phrased in terms of non-contractible loops that begin and end at
infinity, and which would thread through a toroidal horizon. In higher
dimensions,  one can always unlink two loops by moving them apart in a
fourth spatial direction. This suggests that it might be possible for
the event horizon to have non-spherical topology in higher dimensions.
Indeed, in \cite{cai:01}, it was argued that the horizon of a
time-symmetric black hole in five dimensions must have topology given
by a connected sum of $S^3$ and $S^1 \times S^2$ terms, subject to the
weak energy condition. Nevertheless, no example of an asymptotically
flat solution with a non-spherical event horizon has ever been
found\footnote{ A solution with a regular, though degenerate, horizon
of topology $S^1\times S^2$  has been found in \cite{emparan:01a}.
Although not asymptotically flat, this solution has a spacelike
infinity of topology $S^3$, which distinguishes it from examples in
which horizons of non-spherical topology are constructed by taking
spacelike infinity to have non-spherical topology.}.  One aim of the
present paper is to provide an example of such a spacetime.

We will, for simplicity, consider only the vacuum Einstein equations. A
lot of work has been devoted to finding exact solutions of these
equations in dimensions $D>4$. Most of this work has looked for
solutions with a Kaluza-Klein (KK) interpretation, i.e., solutions with
a Killing vector field along which one can perform dimensional
reduction to get a sensible lower dimensional spacetime. For example,
KK black hole solutions were discussed in \cite{KKblackholes} and a  KK
monopole solution was presented in \cite{KKmonopole}. KK
generalizations of the C-metric \cite{kinnersley:70} and Ernst metric
\cite{ernst:76} were presented in \cite{dowker:94}. Other axially
symmetric solutions in KK theory have been discussed in
\cite{gibbons:87, lee, matosetal, melnikovetal}.

Less work has been devoted to finding solutions of the $D$-dimensional
vacuum Einstein equations that do {\it not} admit a KK interpretation,
either because they do not admit an appropriate Killing vector field
along which KK reduction can be performed, or because the reduced
spacetime has pathological features. Examples of such spacetimes are
provided by higher dimensional versions of the Schwarzschild and Kerr
black holes \cite{tangherlini:63,myers:86}. When Wick rotated, these
solutions {\it do} admit KK interpretations as describing instabilities
of the KK vacuum \cite{witten:82} or of KK magnetic fields
\cite{dowker:95, dowker:96}, but their most natural interpretation is
certainly as higher dimensional black holes.

The purpose of the present paper is to obtain and analyze the higher
dimensional analogues of Weyl's class of solutions. Depending on which
feature of Weyl's class one focuses on, there are several directions in
which one can try to extend it to higher dimensions. One possibility is
to seek the class of $D$-dimensional solutions that are static and
axisymmetric, in the sense that they admit an isometry group ${\bf R}
\times O(D-2)$ (with ${\bf R}$ being time translations). However, this
has been tried before \cite{myers:87} without success. Instead, observe
that Weyl's solutions can be characterized as having two orthogonal
commuting Killing vector fields. Hence an alternative way to generalize
Weyl's solutions to higher dimensions is to find all solutions of the
vacuum Einstein equations that admit $D-2$ orthogonal commuting Killing
vector fields. This is done in section \ref{sec:weyl} of this paper.

As in four dimensions, the higher-dimensional Weyl class of solutions is
parametrized in terms of axisymmetric harmonic functions in an auxiliary
flat space. Actually, there are {\it two} classes of Weyl solutions in
higher dimensions. The first, and the most interesting one, is
parametrized in terms of $D-3$ harmonic functions in three-dimensional
flat space, and is the natural analogue of the $D=4$ Weyl solutions
discussed above. The second class of solutions (discussed in Appendix
\ref{app:special}) is parametrized in terms of $D-4$ harmonic
functions in {\it two}-dimensional flat space, and therefore has no
$D=4$ analogue. 

Although Weyl's construction in $D=4$ describes an infinite class of
solutions, most of them are unphysical in the sense that they are not
asymptotically flat, or have naked curvature singularities on the axis
of symmetry\footnote{E.g., a spherically symmetric point source for $U$
results in a singular, {\it non-spherical} Chazy-Curzon particle.}. The
same is true for $D>4$. In order to select candidate Weyl solutions
that might be of physical importance, recall that for $D=4$, the
harmonic function $U$ can be regarded as a Newtonian potential produced
by an axisymmetric source. It turns out that the most interesting $D=4$
Weyl solutions all have sources of the same form, namely thin
rods on the axis of symmetry. In section \ref{sec:known}, known $D>4$
Weyl solutions of physical importance are analyzed. Their harmonic
functions also always correspond to thin rods on the axis of symmetry
in the auxiliary three-dimensional flat space. 

A natural classification scheme for such solutions is presented in
section \ref{sec:newsol}. In this scheme, the `zeroth' class consists
simply of flat space. The first non-trivial class contains just the
$D=4$ and $D=5$ Schwarzschild solutions (the $D>5$ Schwarzschild
solutions do not admit $D-2$ commuting Killing vector fields and are
therefore not Weyl solutions), and their Wick rotations. These Wick
rotations describe objects known as `KK bubbles'. If one considers the
Euclideanized $D=4$ Schwarzschild solution, then the solution looks
asymptotically like ${\bf R}^3\times S^1$, the $S^1$ corresponding to
Euclidean time, which is periodically identified and can be regarded as
a KK compactified dimension. However, the actual topology of the
solution is ${\bf R}^2\times S^2$. The size of the two-spheres at
constant radius decreases from infinity to a minimum non-zero value at
the location of the Euclidean horizon, where a non-contractible $S^2$
lies. At this point, the KK circles smoothly round off and space cannot
be continued past this radius. By adding a flat Lorentzian time
direction one obtains a solution to $D=5$ KK theory where the
non-contractible sphere is a static `bubble of nothing'. It is known to
be unstable \cite{gross:82}. A related solution is obtained by Wick
rotating both the time and one of the ignorable angular coordinates of
the five-dimensional Schwarzschild solution. The Wick-rotated angle then
becomes a boost coordinate and the solution describes a bubble
exponentially expanding in the five-dimensional KK vacuum
\cite{witten:82}. Its fully Euclideanized version is an instanton
mediating the decay of the KK vacuum $M^{1,3}\times S^1$.

The second class of Weyl solutions contains the $D=4$ C-metric as well
as three new solutions. The most interesting of these is a Wick
rotated version of a $D=5$ metric discussed in \cite{chamblin:97}, and
can also be related to the KK C-metric of \cite{dowker:94}. It is a
static, asymptotically flat solution with an event horizon of topology
$S^1 \times S^2$, i.e., it is a {\it black ring}. 
This is the first example of an
asymptotically flat solution of the vacuum Einstein equations that has
an event horizon of non-spherical topology. The solution is not
entirely satisfactory since it has a conical singularity, but it will
be shown in a separate publication that this singularity can be
eliminated if the ring rotates \cite{emparan:01}.

The two other new solutions in the same class as the black ring and the
C-metric both describe superpositions of black objects with static KK
bubbles. These solutions are entirely regular outside an event horizon.
The first is a $D=5$ solution describing a black hole sitting in the
throat of a static KK bubble. The second is a $D=6$ solution describing
a loop of black string with horizon topology $S^3 \times S^1$ sitting
in the throat of a static KK bubble. These solutions asymptote to,
respectively, the KK vacua $M^{1,3}\times S^1$ and $M^{1,4}\times S^1$.
Both are expected to be unstable. In fact, the evolution of the
instability of the former solution can be obtained by a Wick rotation
of the black ring. If these solutions are Euclideanized then they give
new non-singular instantons for the decay of the $T^2$ compactified KK
vacuum in $D=5$ and $D=6$. 

Many of the solutions we describe are naturally interpreted in terms of
KK compactification along the orbits of one or several of the Killing
vector fields. When there is more than one Killing vector field with
closed orbits, one can often dimensionally reduce along different
linear combinations of them. Physically distinct reduced spacetimes can
therefore arise from the same higher dimensional spacetime. A good
example is the KK C-metric and the KK Ernst solution, which are locally
isometric in five dimensions \cite{dowker:95}. With this in mind,
different KK reductions of the new solutions found in this paper are
briefly discussed in section \ref{sec:newKK}.

Multi-black hole configurations can be readily constructed within Weyl's
class, and are briefly discussed in section \ref{sec:multibh}.
Finally, Section \ref{sec:discussion} contains the conclusions of this
work.

\sect{Generalized Weyl solutions}

\label{sec:weyl}

\subsection{Integrable submanifolds}

\label{sec:integrable}

The first step in generalizing Weyl's construction to more than four
dimensions is to find a convenient coordinate chart for the  general
$D$-dimensional line element admitting $D-2$ commuting Killing  vector
fields (orthogonality of these vector fields will not be assumed yet).
This is a simple generalization of what is done in four dimensions (see
\cite{wald:84} for a review). It will be assumed that the metric is
Riemannian or Lorentzian. Let $\xi_{(i)}$ denote the Killing vector
fields, $1\le i \le D-2$. Since these commute, it is possible to choose
coordinates $(x^i,y^1,y^2)$ such that $\xi_{(i)} = \partial/\partial
x^i$ with the metric coefficients depending only on $y^1$ and $y^2$.

The next step is to show that one can choose the coordinates $y^1$ and
$y^2$ to span two-dimensional surfaces orthogonal to all of the
$\xi_{(i)}$. In order to do this, one has to show that the
two-dimensional subspaces of the tangent space orthogonal to all of the
vectors $\xi_{(i)}$ are integrable, i.e., tangent to two-dimensional
surfaces. Sufficient conditions for integrability are supplied by the
following theorem:

{\bf Theorem}. Let $\xi_{(i)}$, $1 \le i \le D-2$ be commuting Killing
vector fields such that for each $i$, (a) $\xi_{(1)}^{[\mu_1} \xi_{(2)}^{\mu_2}
\ldots \xi_{(D-2)}^{\mu_{D-2}} \nabla^{\nu} \xi_{(i)}^{\rho]}$
vanishes at at least one point of the spacetime (not necessarily the
same point for every $i$), and
(b) $\xi_{(i)}^{\nu} R_{\nu}^{[\rho} \xi_{(1)}^{\mu_1}
\xi_{(2)}^{\mu_2} \ldots \xi_{(D-2)}^{\mu_{D-2}]} = 0$. 
Then the two planes orthogonal to the $\xi_{(i)}$ are integrable. 

The proof of this theorem is a straightforward generalization of the
corresponding theorem in four dimensions, as given in
\cite{wald:84}. In this paper, only
vacuum solutions of the Einstein equations will be considered so
condition (b) is trivially satisfied. Condition (a) is less obvious;
in four dimensions it is usually assumed that one of the Killing
vector fields is an angular coordinate corresponding to rotations
about an axis of symmetry, and must therefore vanish on this axis,
which ensures that condition (a) is obeyed. The same assumption can be
used to motivate condition (a) in the higher dimensional case. Of
course, this is not the only way in which condition (a) can be
satisfied, so this theorem has wider applicability than just metrics
with an axis of rotational symmetry. 

If the conditions of this theorem are met then the coordinates $y^1$
and $y^2$ can be chosen in one of the orthogonal surfaces and then
extended along the integral curves of the Killing vector fields\footnote{It is
necessary to assume that the $\xi_{(i)}$ are non-null at this point.}.
In this coordinate system, the vectors $\partial /\partial y^i$ are
orthogonal to $\partial/\partial x^j$. If it is further assumed that
the Killing vector fields are orthogonal to each other 
then the metric must take the form
\be
 ds^2 = \sum_{i=1}^{D-2} \epsilon_i e^{2U_i} (dx^i)^2 + g_{ab} dy^a dy^b,
\ee
where $a$ and $b$ take the values $1,2$, the metric coefficients are
independent of $x^i$, and $\epsilon_i = \pm 1$ according to whether
$\xi_{(i)}$ is spacelike or timelike. 

The final step is to use the freedom to perform coordinate
transformations on $y^a$. Locally it is always possible to choose
coordinates such that
\be
 g_{ab} dy^a dy^b = e^{2C} dZ d\bar{Z},
\ee
where $Z$ and $\bar{Z}$ are complex conjugate coordinates if the
transverse space is spacelike, and independent real coordinates if it is
timelike.\footnote{ The term ``Weyl solution'' is usually reserved for
static solutions (i.e. a spacelike transverse space)
but we adopt a more general usage here.} The function
$C$ is independent of $x^i$.

\subsection{Solving the Einstein equations}
\label{sec:solving}

We have shown that any $D$-dimensional metric that admits $D-2$
orthogonal commuting Killing vector fields can be written locally in
the form
\be
\label{eqn:lineelement}
 ds^2 = \sum_{i=1}^{D-2} \epsilon_i e^{2U_i} (dx^i)^2 + e^{2C} dZ
 d\bar{Z},
\ee
where $U_i$ and $C$ are functions of $Z$ and $\bar{Z}$ only, and
$\epsilon_i = \pm 1$. The summation convention will not be used for
indices $i,j,\ldots$.

The components of the curvature tensors of this line element are
calculated in Appendix \ref{app:curvature}. The vacuum Einstein
equations read $R_{\mu\nu} = 0$. The $ij$ component gives
\be
\label{eqn:Einsteinij}
 \partial_Z \left[ \exp\left( \sum_j U_j \right) \partial_{\bar{Z}}
 U_i \right] + \partial_{\bar{Z}} \left[ \exp \left( \sum_j U_j
 \right) \partial_Z U_i \right] = 0.
\ee
Summing this equation over $i$ yields
\be
 \partial_Z \partial_{\bar{Z}} \exp \left( \sum_j U_j \right) = 0,
\ee
which has the general solution
\be
\label{eqn:wdef}
 \sum_j U_j  = \log \left( w(Z) + \tilde{w}(\bar{Z}) \right),
\ee
where $\tilde{w} = \bar{w}$ if $Z$ and $\bar{Z}$ are complex
conjugate, but $w$ and $\tilde{w}$ are independent real functions if
$Z$ and $\bar{Z}$ are real coordinates. Substituting equation
\ref{eqn:wdef} into equation \ref{eqn:Einsteinij} yields
\be
 \label{eqn:Aieq}
 2 ( w + \tilde{w} ) \partial_Z \partial_{\bar{Z}} U_i + \partial_Z w
 \partial_{\bar{Z}} U_i + \partial_{\bar{Z}} \tilde{w} \partial_Z U_i
 = 0.
\ee

If $w$ is non-constant then $R_{ZZ}=0$ can be rearranged to give
\be
 \partial_Z C = \frac{ \sum_i \partial^2_Z U_i}{\sum_i \partial_Z U_i} +
 \frac{1}{2} \sum_i \partial_Z U_i - \frac{ \sum_{i<j} \partial_Z U_i
 \partial_Z U_j }{ 2 \sum_i \partial_Z U_i}.
\ee
A similar equation arises from $R_{\bar{Z} \bar{Z}} = 0$ (assuming
that $\tilde{w}$ is non-constant):
\be
 \partial_{\bar{Z}} C = \frac{ \sum_i \partial^2_{\bar{Z}} U_i}{\sum_i
 \partial_{\bar{Z}}  U_i} +
 \frac{1}{2} \sum_i \partial_{\bar{Z}} U_i - \frac{ \sum_{i<j}
 \partial_{\bar{Z}}  U_i \partial_{\bar{Z}} U_j }{ 2 \sum_i
 \partial_{\bar{Z}} U_i}.
\ee
The first two terms of these equations be integrated immediately, 
using equation \ref{eqn:wdef} to give
\be
\label{eqn:Csol}
 C = \frac{1}{2} \log \left(\partial_Z w \partial_{\bar{Z}} \tilde{w}
 \right)  + \nu,
\ee
where
\be
\label{eqn:dZnu}
 \partial_Z \nu = - \frac{w + \tilde{w}}{\partial_Z w} \sum_{i<j}
 \partial_Z U_i \partial_Z U_j,
\ee 
\be
\label{eqn:dZbarnu}
 \partial_{\bar{Z}} \nu = - \frac{w + \tilde{w}}{\partial_{\bar{Z}} \tilde{w}} 
 \sum_{i<j} \partial_{\bar{Z}} U_i \partial_{\bar{Z}} U_j.
\ee 
The integrability condition for $\nu$ is
\be
 \partial_Z \partial_{\bar{Z}} \nu = \partial_{\bar{Z}} \partial_Z
\nu.
\ee
It is straightforward to check that this equation is indeed satisfied
by using equations \ref{eqn:wdef} and \ref{eqn:Aieq}. These equations 
also ensure that the remaining Einstein equation $R_{Z \bar{Z}} = 0$
is satisfied.

The only assumptions made above were that $w(Z)$ and
$\tilde{w}(\bar{Z})$ are non-constant. The special cases when one (or
both) of these functions is constant will be dealt with in Section
\ref{sec:special}. With this exception, it has been demonstrated that
the most general solution of the $D$-dimensional Einstein equations that
admits $D-2$ orthogonal commuting Killing vector fields takes the form
\ref{eqn:lineelement}, where $U_i$ are solutions of \ref{eqn:Aieq}
subject to the constraint \ref{eqn:wdef}, and $C$ is given by equation
\ref{eqn:Csol}. The function $\nu$ in this equation is given by
integrating equations \ref{eqn:dZnu} and \ref{eqn:dZbarnu}. 

The constraint \ref{eqn:wdef} can be eliminated by using
it to express, say, $U_1$ in terms of $U_2 \ldots
U_{D-2}$. If this is done then $C$ can be written
\be
 C = \frac{1}{2} \log \left(\partial_Z w \partial_{\bar{Z}} \tilde{w}
 \right) - \sum_{i>1} U_i + \gamma,
\ee
where $\gamma$ is given by integrating
\be
\label{eqn:dZgamma}
 \partial_Z \gamma = \frac{w + \tilde{w}}{\partial_Z w} \left[
 \sum_{i>1} \left( \partial_Z U_i \right)^2 + \sum_{1<i<j} \partial_Z
 U_i \partial_Z U_j \right],
\ee
\be
\label{eqn:dZbargamma}
 \partial_{\bar{Z}} \gamma = \frac{w + \tilde{w}}{\partial_{\bar{Z}}
 \tilde{w}} \left[ \sum_{i>1} \left( \partial_{\bar{Z}} U_i \right)^2 +
 \sum_{1<i<j} \partial_{\bar{Z}} U_i \partial_{\bar{Z}} U_j \right].
\ee

\subsection{Relation to Laplace's equation}

\label{sec:laplace}

Since $w$ and $\tilde{w}$ have been assumed non-constant, it is
legitimate to perform a coordinate transformation from $Z$ and $\bar{Z}$
to $w(Z)$ and $\tilde{w}(\bar{Z})$. In four dimensions, these are
referred to as ``Weyl's canonical coordinates'' \cite{kramer:80}.
This gives
\be
\label{eqn:weylmetric}
 ds^2 = \sum_i \epsilon_i e^{2U_i} (dx^i)^2 + e^{2\nu} dw d\tilde{w}. 
\ee
This coordinate transformation is conformal. Equations
\ref{eqn:Aieq}, \ref{eqn:dZnu} and \ref{eqn:dZbarnu}
are conformally invariant so the transformation just replaces
$\partial_Z$ by $\partial \equiv \partial_w$ and $\partial_{\bar{Z}}$
by $\bar{\partial} \equiv \partial_{\tilde{w}}$. Then the solution is
determined by the following equations
\be
 \label{eqn:wdef2}
 \sum_i U_i = \log ( w + \tilde{w}),
\ee
\be
 \label{eqn:Aieq2}
 2 (w + \tilde{w}) \partial \bar{\partial} U_i + \partial U_i +
 \bar{\partial} U_i = 0,
\ee
\be
 \label{eqn:dwnu}
 \partial \nu = - (w+ \tilde{w}) \sum_{i<j} \partial U_i \partial U_j,
\ee
\be
 \label{eqn:dwbarnu}
 \bar{\partial} \nu = - (w + \tilde{w}) \sum_{i<j} \bar{\partial} U_i
 \bar{\partial} U_j.
\ee
If one prefers to eliminate the constraint \ref{eqn:wdef2} then the
metric takes the form
\be
\label{eqn:metric1}
 ds^2 = \exp \left( -2\sum_{i>1} U_i \right)\left[
 e^{2\gamma}dw d\tilde{w} + \epsilon_1 (w+\tilde{w})^2 (dx^1)^2 \right]
 + \sum_{i>1} \epsilon_i e^{2U_i} (dx^i)^2,
\ee
with $\gamma$ determined by
\be
\label{eqn:dwgamma}
 \partial \gamma = (w + \tilde{w}) \left[
 \sum_{i>1} \left( \partial U_i \right)^2 + \sum_{1<i<j} \partial
 U_i \partial U_j \right],
\ee
\be
 \label{eqn:dwbargamma}
 \bar{\partial} \gamma = (w + \tilde{w}) \left[\sum_{i>1} \left(
 \bar{\partial} U_i \right)^2 + \sum_{1<i<j} \bar{\partial} U_i
 \bar{\partial}  U_j \right].
\ee 
If $Z$ and $\bar{Z}$ are complex conjugate coordinates then, as
mentioned above, one must take $\tilde{w} = \bar{w}$. Introduce real
coordinates $(r,z)$ by $w = r + iz$, so the canonical form of the metric
is
\be
 ds^2= \sum_i \epsilon_i e^{2U_i} (dx^i)^2 + e^{2\nu}(dr^2+dz^2). 
\ee
Equation \ref{eqn:Aieq2} then
takes the form
\be
 \frac{\partial^2 U_i}{\partial r^2} + \frac{1}{r}\frac{\partial
 U_i}{\partial r} + \frac{\partial^2 U_i}{\partial z^2} = 0,
\ee
which is just Laplace's equation in three-dimensional flat space with
metric 
\be
 ds^2 = dr^2 + r^2 d\theta^2 + dz^2.
\ee
The function $U_i$ is independent of the (unphysical) coordinate
$\theta$, i.e., it is axisymmetric. The solution is therefore
specified by $D-3$ independent axisymmetric solutions of Laplace's
equation in three-dimensional flat space. There are only $D-3$
independent $U_i$ because of the constraint \ref{eqn:wdef2}, which can
now be written
\be
\label{eqn:Acons} 
 \sum_i U_i = \log r + {\rm constant},
\ee
where the constant term can be freely adjusted by rescaling the
coordinates $x^i$. Note that $\log r$ is the solution of Laplace's
equation that describes the Newtonian potential produced by an infinite
rod of zero thickness lying along the $z$-axis, with constant mass
$1/2$ per unit length (in units $G=1$).  The solutions for $U_i$ can
also be thought of as Newtonian potentials produced by certain sources,
so the constraint \ref{eqn:Acons} states that these sources must add up
to give an infinite rod. Note that the solution is completely
determined by these sources. The sources for $U_i$ will sometimes be
referred to as the sources for $x^i$.

For $D=4$, the metric \ref{eqn:metric1} can be brought to the standard
form of equation \ref{eqn:4dweylmetric} by taking $\epsilon_1 =
-\epsilon_2 = 1$, $x^2 = t$, $x^1 = \phi/2$ and $U_2=U$. 
However, this form obscures the symmetry between $x^1$ and $x^2$, and
hides the fact that solutions which have different sources for $U$ may
actually be equivalent under interchange of $t$ and $\phi$. We will
illustrate this point with an example in section \ref{sec:schwarz}.

If $w$ and $\tilde{w}$ are real coordinates then they can be viewed as
advanced and retarded null coordinates. Introduce new coordinates
$(t,r)$ defined by $w = r+t$, $\tilde{w} = r-t$. Then equation
\ref{eqn:wdef2} becomes
 \be
 -\frac{\partial^2 U_i}{\partial t^2} + \frac{\partial^2 U_i}{\partial
 r^2} + \frac{1}{r} \frac{\partial U_i}{\partial r} = 0,
\ee
which is just the wave equation in a three-dimensional flat spacetime
with metric
\be
 ds^2 = -dt^2 + dr^2 + r^2 d\theta^2,
\ee
with the function $U_i$ independent of the unphysical coordinate
$\theta$. The solution is specified by $D-2$ axisymmetric solutions of
the wave equation in three-dimensional flat spacetime but only $D-3$
of these are independent because the constraint \ref{eqn:wdef2} states
that these solutions must add up to the static solution describing a point
source at the origin of polar coordinates.

\subsection{Special classes of solutions}

\label{sec:special}

It was assumed above that $w(Z)$ and $\tilde{w}(\bar{Z})$ are
non-constant, but there is the possibility that one or both of these
quantities is constant. These solutions are discussed in Appendix
\ref{app:special}. They will be referred to as {\it special Weyl
solutions} to distinguish them from those of the previous sections,
which will be referred to as {\it generic Weyl solutions}. For $D=4$ the
special solutions are either flat space or pp-waves. The interpretation
of the solutions in $D>4$ is unclear, so it might be interesting to
investigate them further. In $D=5$, most of the static special solutions 
appear to be nakedly singular but there may be exceptions.

\bigskip

In the rest of the paper we will consider only static ($\tilde{w} =
\bar{w}$) Weyl solutions of the generic class.

\sect{Weyl form of known solutions}

\label{sec:known}

Generic Weyl metrics are characterized by axisymmetric harmonic
functions in three-dimensional flat space. If such functions are
regarded as Newtonian potentials produced by axisymmetric sources, then
Weyl solutions can be completely characterized by these sources. In
order to identify the types of sources that might be relevant in
attempting to find interesting new solutions, we study the Weyl form for
some known physically relevant metrics.

\subsection{Flat space}

\label{sec:flat}

We study in Appendix \ref{app:flat} the circumstances under which the
metric \ref{eqn:lineelement} is flat. There are three possibilities.
This first is trivially given by taking all of the functions $U_i$ to be
constant\footnote{This can be regarded as belonging to the special Weyl
class.}. The second possibility corresponds to all
but one of the functions $U_i$ being constant, and one of the $U_i$
(say, $U_1$) being the potential of an infinite rod along the
$z$-axis, $U_1=\log r+{\rm constant}$. The metric in this case can be
brought to the form (see Appendix \ref{app:flat})
\be
 ds^2 = \epsilon_1 \xi^2 (dx^1)^2 + \sum_{i=2}^{D-2}
 \epsilon_i (dx^i)^2 + d\xi^2 + d\eta^2,
\ee
so if $\epsilon_1 = +1$ then $x^1$ is an azimuthal angle and if
$\epsilon_1 = -1$ then $x^1$ is a Rindler (boost) time coordinate.

\begin{figure}
\centering{\psfig{file=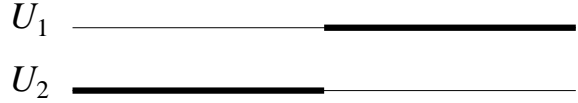,width=3.in}}
\caption{Sources for the harmonic functions of one of the Weyl forms
of flat space. The thin lines denote the $z$-axis and the thick lines
denote thin rods along this axis. The left and right ends of
the figure are to be interpreted as at $z=-\infty$ and $z=+\infty$
respectively. The sources for $U_1$ and $U_2$ are
semi-infinite rods of mass $1/2$ per unit length. The $U_1$
source lies along $z \ge a$ and the $U_2$ source along $z \le a$ for
some $a$.
In the classification of Section \ref{sec:classes}, this is a
class 0 solution.}
\label{fig:sources0}
\end{figure}

The third possibility corresponds to all but two of
the functions $U_i$ being constant, with one of the remaining two being
the potential of a semi-infinite rod along the $z\ge a$ portion of the
$z$-axis (for some $a$), and the other being the potential of a
semi-infinite rod along the $z \le a$ portion of the $z$-axis (see
figure \ref{fig:sources0}). 

In terms of $w$,
\be
 U_1 = \log | \Re \sqrt{a \pm iw} | + {\rm constant},
\ee
\be 
 U_2 = \log | \Re \sqrt{-a \mp iw}| + {\rm constant}.
\ee
Writing these in terms of $r,z$ gives
\be
 U_1 = \frac{1}{2} \log \left[a \mp z + \sqrt{(a\mp z)^2 + r^2}
\right] + {\rm constant},
\ee
\be
 U_2 = \frac{1}{2} \log \left[-a \pm z + \sqrt{ (-a \pm z)^2 + r^2}
\right] + {\rm constant}.
\ee
The upper sign choice corresponds to $U_1$ being the potential of a
semi-infinite rod $z \ge a$ and $U_2$ being that of a semi-infinite rod
$z \le a$. The lower sign choice corresponds to the source for $U_1$
being a semi-infinite rod $z \le -a$ and the source for $U_2$ a
semi-infinite rod $z \ge -a$. The rods are all on the $z$-axis, have
zero thickness, and mass $1/2$ per unit length. The metric in this
case can be brought to the form (see Appendix \ref{app:flat})
\be
 ds^2 = \epsilon_1 \xi^2 (dx^1)^2 + \epsilon_2 \eta^2 (dx^2)^2 +
\sum_{i=3}^{D-2} \epsilon_i (dx^i)^2 + d\xi^2 + d\eta^2,
\label{eqn:flat}
\ee
so the coordinates $x^1$ and $x^2$ are azimuthal angles or Rindler
time coordinates according to the signs of $\epsilon_1$ and
$\epsilon_2$. 

\subsection{The Schwarzschild solution}

\label{sec:schwarz}

The $D$-dimensional Schwarzschild solution has isometry group ${\bf R}
\times O(D-1)$. To write it in Weyl form, $D-2$ orthogonal commuting
Killing vector fields are required. For the Schwarzschild solution,
this occurs only for $D=4,5$. Hence only the four and five-dimensional
Schwarzschild solutions can be written in Weyl form. The Weyl form of
the four-dimensional Schwarzschild solution is well known so it will
be discussed only briefly here. 

\begin{figure}
\centering{\psfig{file=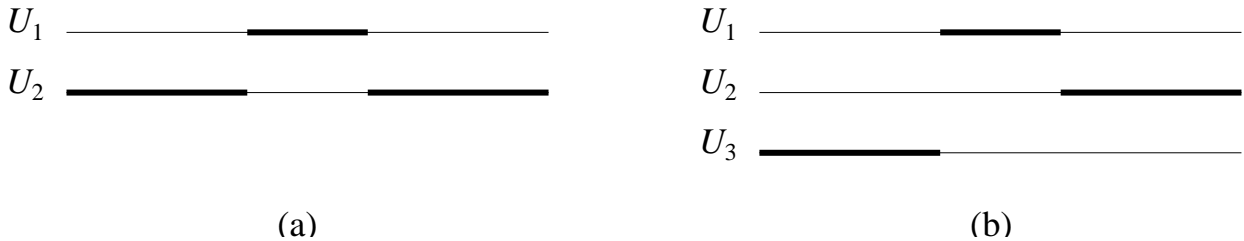,width=6.5in}}
\caption{Sources for (a) the four dimensional and (b) five dimensional
Schwarzschild solutions. The black hole interpretation requires that
$x^1$ is the timelike coordinate. If in (a) $x^2$ is the timelike
coordinate then this describes an expanding bubble in the $M^{1,2}\times
S^1$ vacuum. If both $x^1$ and $x^2$ are spacelike then this describes a
static KK $S^2$ bubble (when a trivial time direction is added). 
If in (b) $x^2$ (or $x^3$) corresponds to time,
then it describes an expanding bubble in the $M^{1,3}\times S^1$ vacuum.
If $x^1$, $x^2$ and $x^3$ are all spatial coordinates, then it describes
an $S^3$ bubble. In the classification of Section \ref{sec:classes},
these solutions are class I.}
\label{fig:sources1}
\end{figure}

For $D=4$, a generic Weyl solution can be converted to the form of
equation \ref{eqn:4dweylmetric} as described in section
\ref{sec:laplace}. For the Schwarzschild metric, 
the function $U$ is given by
\be
 U = -\frac{1}{2} \log \left[\frac{M-z + \sqrt{(M-z)^2+r^2}}{-M-z +
\sqrt{(M+z)^2 + r^2}} \right],
\ee
where $M$ is the Schwarzschild mass parameter. $U \equiv 
U_1$ is the potential of
a finite rod along the $-M \le z \le M$ portion\footnote{
One is free to shift the rod to any position on the
$z$-axis with a transformation $z\rightarrow z + a$.}
of the $z$-axis. The
rod has vanishing thickness and mass $1/2$ per unit length. It follows
from the constraint \ref{eqn:Acons} that the function $U_2$ 
must be the potential produced by semi-infinite
rods $z \ge M$ and $z \le -M$. These sources are depicted in figure
\ref{fig:sources1}(a).

Note that our approach makes clear the nature of the solution of the
$D=4$ Weyl class \ref{eqn:4dweylmetric} that is obtained by taking two
semi-infinite rod sources for $U$. In our approach it simply
corresponds to interchanging $x^1$ and $x^2$, i.e, interchanging the
time $t$ and azimuthal angle $\phi$ coordinates. This gives a
four-dimensional analogue of the expanding KK bubble of
\cite{witten:82}, and describes the decay of the $D=4$ KK vacuum
$M^{1,2}\times S^1$ \cite{mann:95}.

The five-dimensional Schwarzschild metric can be written in
Schwarzschild coordinates as
\be
 ds^2 = -\left( 1-\frac{\mu}{R^2} \right) dt^2 + 
 \left( 1-\frac{\mu}{R^2} \right)^{-1} dR^2 + R^2 d\theta^2 + R^2
 \sin^2 \theta d\phi^2 + R^2 \cos^2 \theta d\psi^2,
\label{eqn:5dsch}
\ee
where $0 \le \theta \le \pi/2$, and $\phi \sim \phi+2\pi$, $\psi \sim
\psi+2\pi$. There are clearly three orthogonal commuting Killing
vector fields. Take $x^1 = t$, $x^2 = \phi$, $x^3 = \psi$ with
$\epsilon_1 = -1$, $\epsilon_2 = \epsilon_3 = 1$. Then one can read off
\be
e^{U_1} = \left( 1-\frac{\mu}{R^2} \right)^{1/2}, \qquad e^{U_2} = R
\sin
\theta, \qquad e^{U_3} = R \cos \theta. 
\ee 
The constraint \ref{eqn:Acons} gives
\be
\label{eqn:UABcons}
 U_1+U_2+U_3 = \log r,
\ee
where the constant term has been absorbed into the normalization of
$r$. This equation implies
\be
\label{eqn:rschw}
 r = \frac{1}{2} \left(1 - \frac{\mu}{R^2} \right)^{1/2} R^2 \sin 2
 \theta.
\ee
To bring the metric to Weyl form, it is necessary to define
$z$ such that
\be
 dr^2 + dz^2 \equiv dw d\bar{w} \propto 
 \left( 1-\frac{\mu}{R^2} \right)^{-1} dR^2 + R^2
 d\theta^2.
\ee
Substituting the ansatz $z = g(R) \cos 2 \theta$ into this equation
then uniquely determines $g(R)$, giving
\be
\label{eqn:zschw}
 z = \frac{1}{2} \left(1 - \frac{\mu}{2R^2} \right) R^2 \cos 2 \theta.
\ee
It also possible to read off $\gamma$:
\be
 e^{2\gamma} = \frac{1}{4} \left( 1 - \frac{\mu}{R^2} +
\frac{\mu^2}{4R^4} \cos^2 2\theta \right)^{-1} R^2 \sin^2 2\theta.
\ee
It remains to write the functions $U_i$ in terms of $r$ and $z$. To do this,
let $X = e^{U_2}$ and $Y=e^{U_3}$. Equations \ref{eqn:rschw} and
\ref{eqn:zschw} can then be written as
\be
 r^2 = \left(1-\frac{\mu}{X^2+Y^2} \right) X^2 Y^2, \qquad z =
 \frac{1}{2} \left( 1 - \frac{\mu}{2(X^2+Y^2)} \right) (Y^2-X^2),
\ee
which can be rearranged to give
\be
 2 Y^4 - (\mu + 4 z) Y^2 - 2 r^2 = 0,
\ee
\be
 2 X^4 - (\mu - 4 z) X^2 - 2 r^2 = 0.
\ee
Solving these yields
\be
 U_2 = \frac{1}{2} \log \left[\frac{\mu}{4} - z + \sqrt{
 \left(\frac{\mu}{4} - z \right)^2 + r^2 } \right],
\ee
\be
 U_3 = \frac{1}{2} \log \left[\frac{\mu}{4} + z + \sqrt{
 \left(\frac{\mu}{4} + z \right)^2 + r^2 } \right].
\ee
$\gamma$ can be written in terms of $w$ and $\bar{w}$ to check
that equations \ref{eqn:dwgamma} and \ref{eqn:dwbargamma} are obeyed.
The explicit expression will not be written out here
since it can be obtained as a special case
of more general expressions given later in this paper.
$U_2$ is the potential of a semi-infinite rod with vanishing thickness
and mass $1/2$ per unit length positioned 
along the $z$-axis at $z \ge \mu/4$. $U_3$ is the potential of an
identical rod along the $z$-axis at $z \le -\mu/4$.  
The function $U_1$ is obtained from
\ref{eqn:UABcons} and is the potential of a rod along the $-\mu/4 \le z
\le \mu/4$ portion of the $z$-axis, again with vanishing thickness and
mass $1/2$ per unit length. See figure \ref{fig:sources1}(b). Note that
the source corresponding to the time coordinate is a finite rod for both
the $D=4$ and the $D=5$ Schwarzschild solutions. 

{\bf Black branes.}
For $D>5$, the $D$-dimensional Schwarzschild solution is not a
generalized Weyl solution. However, the black branes obtained by
taking products of the $D=4$ or $D=5$ Schwarzschild solution with flat
space are easily seen to be Weyl solutions. The functions $U_i$
associated with the flat directions are all constant, and those
associated with the Schwarzschild directions can be read off from
the results of this section.

\subsection{Other four-dimensional solutions}

Other physically relevant four-dimensional Weyl solutions are:

{\bf The Israel-Khan solutions} \cite{israel:64}.
These describe finitely many collinear black holes in static
equilibrium. The forces holding them apart arise from conical deficits
in the form of struts between the black holes, or cosmic strings
extending to infinity. If the metric is written in the Weyl form
\ref{eqn:4dweylmetric} then the sources for $U$ are finite rods along the
$z$-axis. The rods have zero thickness, mass $1/2$ per unit length,
and do not intersect. The length of each rod determines the mass
of the corresponding black hole. If one considers infinitely many such
rods of equal length and equally spaced 
then one can eliminate the need for conical singularities and
obtain a solution describing an infinite line of black holes
\cite{myers:87}. 

\begin{figure}
\centering{\psfig{file=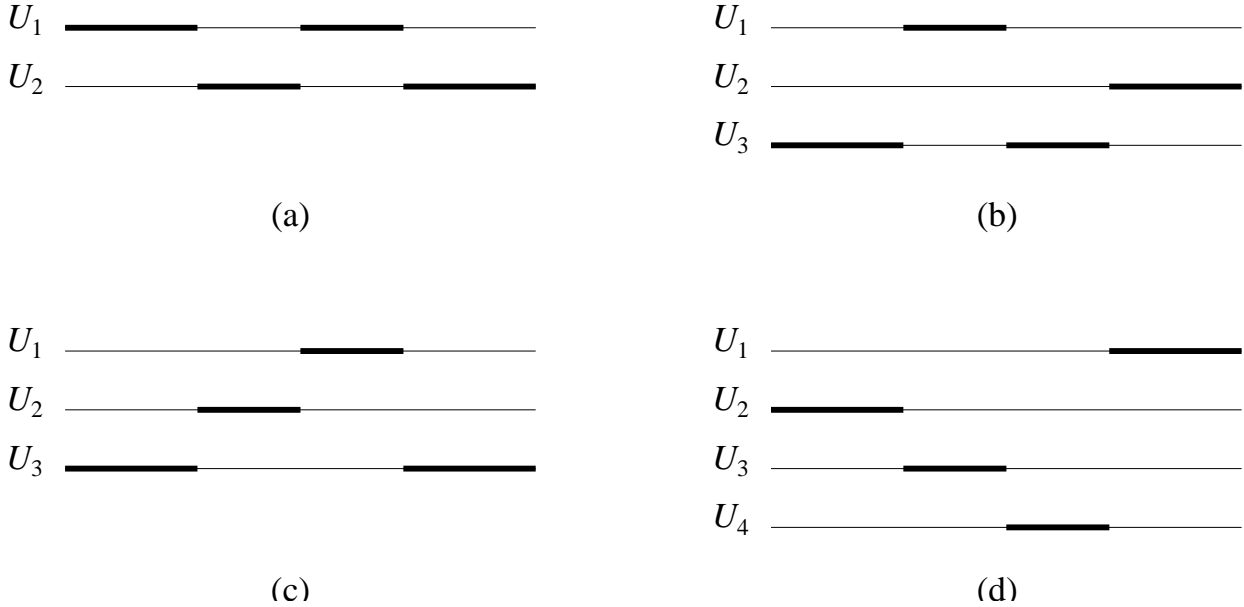,width=6.5in}}
\caption{Sources for (a) C-metric; (b) black ring; (c) black hole plus
KK bubble; (d) black string and KK bubble. Note that the sources for the
$U_i$'s have to add up to an infinite rod. In the classification of
Section \ref{sec:classes}, these solutions are class II.}
\label{fig:sources2}
\end{figure}

{\bf The C-metric} \cite{kinnersley:70}. 
The C-metric is a four-dimensional metric that describes two black
holes accelerating apart. The force for the acceleration is provided by
a conical deficit, which occurs either in the form of a strut between
the two black holes or as a cosmic string stretching off to infinity
from each hole. The Weyl form of the C-metric was obtained in
\cite{godfrey:72, bonnor:83}. The function $U$ is the potential of a
finite rod and a semi-infinite rod, which do not intersect. Both rods
lie along the $z$-axis, have zero thickness and have mass $1/2$ per
unit length (see figure \ref{fig:sources2}(a) with $U_1 = U$). 
The finite rod corresponds
to one of the black holes and the semi-infinite rod is responsible for
the acceleration field (the second black hole lies beyond an
acceleration horizon, so it is not apparent in the Weyl coordinates).
Adding further finite rods sources to $U$ results in a metric
describing multiple accelerating black holes connected by conical
deficits \cite{dowker:01}.

\sect{New solutions}

\label{sec:newsol}

The sources for the solutions discussed in the previous section are all
rods of zero thickness and mass $1/2$ per unit length. The rods are on
the $z$-axis and can be finite, semi-infinite or infinite. More general
sources typically give rise to naked curvature singularities on the axis
of symmetry. Hence, in attempting to find interesting new Weyl solutions
we will consider only sources of this form. The examples of the previous
section allow us to make some general observations that are useful
when analyzing a solution given its set of sources. 

\subsection{General comments}

\label{sec:interpret}

The constraint \ref{eqn:Acons} is very restrictive: It states that the
sources for the different $U_i$'s must add up to an infinite rod along
the $z$-axis, again with mass $1/2$ per unit length. Assuming that only
finitely many rods are present, it follows that either one of the
$U_i$'s has semi-infinite rod sources which extend to $z = \infty$ and
$z=-\infty$, or there is one $U_i$ with a semi-infinite rod source which
extend to $z=\infty$ and another with a semi-infinite rod source which
extend to $z=-\infty$. All of the other $U_i$'s must have bounded
sources consisting of a finite number of finite rods. 

If the source for $U_i$ is bounded (i.e. involves only finitely many
finite rods) then $U_i$ must approach a constant far from the source. It
follows that $x^i$ must be a flat direction in the asymptotic metric. An
example of this is provided by the Schwarzschild metrics, for which the
source corresponding to the time coordinate is a finite rod, and the
time direction does indeed become flat in the asymptotic region. 

Now consider the behaviour near the sources. Assume first that $x^i$ is
a time coordinate. For the $D=4,5$ Schwarzschild solutions,  the source
for $U_i$ is a finite rod and  the region near this source corresponds
to the event horizon of the black hole. This is also true for the
finite rod sources in the C-metric and Israel-Khan solutions. The
semi-infinite rod source of the C-metric corresponds to a horizon that
extends to asymptotic infinity -- this is an acceleration horizon,
which arises because the time coordinate behaves like a boost at
asymptotic infinity.  It was shown above that flat space can also be
written in a Weyl form in which the time coordinate has a semi-infinite
rod source. This source also corresponds to an acceleration horizon,
arising because the time coordinate is a boost, i.e., the Rindler time
coordinate.  To summarize, 
\begin{itemize}
\item finite rod sources for the time
coordinate correspond to event horizons in spacetime, and semi-infinite
rod sources correspond to acceleration horizons.
\end{itemize}
The case in which $x^i$ is a spatial coordinate can be understood by
Euclideanizing some of the metrics discussed above. For the Euclidean
Schwarzschild solution, the finite rod source corresponds to the
``bolt'' where the Euclidean time direction closes off smoothly,
provided it is identified with a suitable period. A similar
interpretation holds for Euclideanized Rindler space, with the only
difference being that in the former case, the bolt is finite in extent
(it is an $S^2$ in $D=4$) whereas in the latter case it extends to
infinity  (it is $R^2$ in $D=4$). These features also occur for the
other examples above. In conclusion, 
\begin{itemize}
\item rod sources for a spatial
coordinate $x^i$ correspond to ``bolts'': fixed-point sets of the
orbits of $\partial/\partial x^i$. 
If the source for $x^i$ extends to infinity,
then the bolt will also extend to infinity, corresponding to an axis of
rotational symmetry with $x^i$ acting as the azimuthal angle.
\end{itemize}
In order to avoid a conical
singularity, $x^i$ has to be periodically identified with a particular
period determined by the sources. If there is more than one source then
there will be several bolts, and the appropriate periods for $x^i$ at
each bolt may differ. In this case, conical singularities will result.
This occurs for the C-metric and Israel-Khan solutions as well as those
of \cite{dowker:01}.

A final fact useful when analyzing Weyl solutions immediately follows
from the above discussion.  Let $x^i$ be a spatial coordinate with a
single finite rod source. Then $x^i$ has to be identified with a
certain period in order to avoid a conical singularity at the bolt
corresponding to the source. Moreover, $x^i$ is a flat coordinate in
the asymptotic region. It follows that $x^i$ is most naturally
interpreted as parametrizing a KK circle at infinity. If there is more
than one finite rod source then it might no longer be possible to
remove all conical singularities by identifying $x^i$ but one would
probably still wish to minimize the number of singularities by an
appropriate identification. Hence, 
\begin{itemize}
\item if a spatial coordinate $x^i$ has only finite
rod sources then it can be interpreted as a KK coordinate in the asymptotic
region.
\end{itemize} 
The rod sources, where the KK circle shrinks to zero size, appear as
singularities in the dimensionally reduced description. 

Now, since at most two of the $U_i$'s have sources extending to
infinity, it follows that at least $D-4$ of the $x^i$'s will have
bounded sources. If one of these is the time coordinate then there will
be at least $D-5$ spatial coordinates with bounded sources so the
asymptotic metric will have at least $D-5$ compactified flat
directions. It follows that no $D>5$ Weyl solution can be
asymptotically flat (in the global sense) if it has sources of the form
being considered here.

\subsection{Classification}

\label{sec:classes}

A solution will be said to be of class $n$ if it has $n$ finite rod
sources (as well as a suitable number of infinite, or semi-infinite rod
sources). 

We make no distinction between metrics related by Wick rotation, so any
of the $x^i$ can be chosen as the time direction. The first few classes
are:

{\bf Class 0.} If there are no finite rod sources then the sources
must be either an infinite rod, or two semi-infinite rods (figure
\ref{fig:sources0}). It was
shown above that the metric is flat in both of these cases, so flat
space is the only Class $0$ solution.

{\bf Class I.} In this class, there is a single finite rod so the other
sources must be two semi-infinite rods (figure \ref{fig:sources1}). 
There are two ways that these sources can be distributed amongst the
$U_i$'s. {\bf (a).} $U_1$ has a finite rod source and $U_2$ has both
semi-infinite rod sources. The other $U_i$ are constant. This is the
four-dimensional Schwarzschild solution (times some flat directions if
$D>4$).  {\bf (b).} $U_1$ has a finite rod source, $U_2$ and $U_3$ have
semi-infinite rod sources. The other $U_i$ are constant. This is the
five-dimensional Schwarzschild solution (times some flat directions if
$D>5$). 

{\bf Class II.} The sources are two finite rods $a_3 \le z \le a_2$ and
$a_2 \le z \le a_1$ and two semi-infinite rods $z \ge a_1$ and $z \le
a_3$ (figure \ref{fig:sources2}). There are four ways to distribute these
sources amongst the $U_i$'s. Flat dimensions (corresponding to constant
$U_i$) will be neglected. {\bf (a).} $U_1$ has a semi-infinite rod
source and a finite rod source, as does $U_2$. This gives the
four-dimensional C-metric. {\bf (b).} $U_1$ has a finite rod source,
$U_2$ has a semi-infinite rod, $U_3$ has a finite rod and a
semi-infinite rod. This is a new ``black ring'' solution that will be
discussed in section \ref{sec:ring}. {\bf (c).} $U_1$ and $U_2$ have the
finite rods as sources, and $U_3$ has both semi-infinite rods as its
sources. This is a new $D=5$ solution describing a superposition of a
black hole with a Kaluza-Klein bubble. {\bf (d).} $U_1$ and $U_2$ have
the semi-infinite rods as sources, $U_3$ and $U_4$ have the finite rods
as sources. This is a new $D=6$ solution describing the superposition of
a black string with a Kaluza-Klein bubble. Solutions (c) and (d) will be
discussed in sections \ref{sec:holebubble} and \ref{sec:stringbubble}
respectively.

The $D=4$ class $n$ solutions for $n>2$ have all been discussed
before. If $n$ is odd, $n = 2k-1$, then the solution is an Israel-Khan
solution describing $k$ black holes. If $n$ is even, $n=2k$ then the
solution is a generalization of the C-metric of the form discussed in
\cite{dowker:01} and describes $k$ accelerating black holes on each
side of an acceleration horizon. Conical singularities are present in 
both cases.

A solution in a given class can be reduced to a solution of a lower
class by either contracting to zero size or expanding to infinity one
of its finite rods. The limits that must be taken to recover a given
solution can be easily deduced by looking at the diagrams for sources
in figures \ref{fig:sources0}, \ref{fig:sources1}, \ref{fig:sources2}.
For example, from fig.~\ref{fig:sources2}(a) we easily see that the
C-metric has a limit where one recovers the $D=4$ Schwarzschild
solution, fig.~\ref{fig:sources1}(a), by taking to infinity the
leftmost endpoint of the $x^2$ rod. Effectively, this amounts to
removing the acceleration horizon from the metric, the well-known limit
where the acceleration of the black hole is set to zero. The C-metric
also has several limits where flat (Rindler) space, figure
\ref{fig:sources0}, is recovered. The II(b) solution that below will be
interpreted as a black ring, fig.~\ref{fig:sources2}(b), similarly
reduces to either a $D=5$ black hole (fig.~\ref{fig:sources1}(b)), or a
black string obtained as the product of the $D=4$ black hole
(fig.~\ref{fig:sources1}(a)) and a flat spatial direction.

\subsection{The black ring}

\label{sec:ring}

In \cite{chamblin:97} an unconventional neutral limit for the KK charged
C-metric (\cite{dowker:94} dualized to have electric charge) was taken.
The resulting metric was interpreted as describing
a pair of KK bubbles being accelerated apart by a conical singularity.
We now show that this metric has a less exotic
interpretation if one Wick rotates it to give
\ba \label{eqn:bring}
 ds^2 &=& -\frac{F(x)}{F(y)} dt^2 \\
 &+& \frac{1}{A^2 (x-y)^2} \left[ F(x)
 \left( (y^2-1) d\psi^2 + \frac{F(y)}{y^2-1} dy^2 \right) + F(y)^2
 \left( \frac{dx^2}{1-x^2} + \frac{1-x^2}{F(x)} d\phi^2 \right)
 \right], \nonumber
\ea
where
\be\label{eqn:fxi}
 F(\xi) = 1-\mu\xi.
\ee
The parameters $\mu$ and $A$ will be taken to lie in the range $0 \le
\mu \le 1$, $A>0$, the coordinate $x$ in the range $-1 \le x \le 1$ and
the coordinate $y$ in the range $y \le -1$. This metric clearly has
three orthogonal commuting Killing vector fields so it is a Weyl
solution. Choosing $t=x^1$, $\psi=x^2$ and $\phi=x^3$, the functions
$U_i$ are given by
\be
 e^{2U_1} = \frac{F(x)}{F(y)},
\ee
\be
 e^{2U_2} = \frac{(y^2-1) F(x)}{A^2 (x-y)^2},
\ee
\be
 e^{2U_3} = \frac{(1-x^2) F(y)^2}{A^2 (x-y)^2 F(x)}.
\ee
In order to identify the sources that produce this solution it is
necessary to work with the coordinates $r,z$. From equation
\ref{eqn:Acons}, it follows that
\be
\label{eqn:rdef}
 r = \frac{\alpha}{A^2 (x-y)^2} \sqrt{F(x)F(y)(1-x^2)(y^2-1)},
\ee
for some positive constant $\alpha$. The coordinate $z$ is obtained
from the requirement
\be
 dr^2 + dz^2 \propto \frac{F(y)}{1-x^2} dx^2 + \frac{F(x)}{y^2-1}
 dy^2,
\ee
which yields
\be
 \frac{\partial z}{\partial x} = \pm \sqrt{ \frac{(y^2-1) F(y)}
 {(1-x^2) F(x)} } \frac{\partial r}{\partial y},
\ee
and 
\be
 \frac{\partial z}{\partial y} = \mp \sqrt{ \frac{ (1-x^2) F(x)}
 {(y^2-1) F(y)} } \frac{\partial r}{\partial x}.
\ee
These equations can be integrated to give
\be
\label{eqn:zdef}
 z = \frac{\alpha (1-xy) (F(x) + F(y))}{2 A^2 (x-y)^2}
\ee
up to a choice of sign and an arbitrary additive constant. 

In order to write the solution in Weyl form, it is convenient to
define
\be 
\label{eqn:aidef}
 a_1 = \alpha/(2A^2), \qquad a_2 = \alpha \mu/(2A^2), \qquad a_3 =
 -\alpha \mu/(2A^2) 
\ee
and then introduce the following notation \cite{israel:64, dowker:01}:
\be
\label{eqn:zidef}
 \zeta_i \equiv z-a_i,
\ee
\be
\label{eqn:Ridef}
 R_i \equiv \sqrt{r^2 + \zeta_i^2},
\ee
\be
\label{eqn:Yijdef}
 Y_{ij} \equiv R_i R_j + \zeta_i \zeta_j + r^2.
\ee
Expressions for these quantities in terms of $x$ and $y$ are given in
Appendix \ref{app:ring}. Using these expressions, it is easily seen 
that if one takes $\alpha=A$ then the Weyl form of the metric is given by
\be
 e^{2U_1} = \frac{R_3 - \zeta_3}{R_2 - \zeta_2},
\ee
\be
 e^{2U_2} = (R_1 - \zeta_1)/A,
\ee
\be
 e^{2U_3} = \frac{(R_1 + \zeta_1) (R_2 - \zeta_2)}{A(R_3 - \zeta_3)},
\ee
\be
 e^{2\nu} = \frac{1+\mu}{4 A } \frac{Y_{23}}{R_1 R_2 R_3} \sqrt{
 \frac{Y_{12}}{Y_{13}} } \sqrt{ \frac{R_2 - \zeta_2}{R_3 - \zeta_3} },
\ee
from which it follows that $U_1$ is the Newtonian potential 
produced by a finite rod
$-\mu/(2A) \le z \le \mu/(2A)$, $U_2$ is the potential produced by
a semi-infinite rod $z \ge 1/(2A)$, and $U_3$ is the potential
produced by a semi-infinite rod $z \le -\mu /(2A)$ and a finite rod
$\mu /(2A) \le z \le 1/(2A)$. Note that, for $\mu=1$, these
sources reduce to those of the five-dimensional Schwarzschild solution
and hence the metric must reduce to the metric of the Schwarzschild
solution, so the function $\nu$ for Schwarzschild can be read off from
the above.

\subsection{Analysis of the black ring}

We now explain why the name ``black ring'' is appropriate by examining
the global structure of this solution.  To start, consider how the
general comments of section \ref{sec:interpret} apply to this solution.
The source for $t$ is a finite rod, so the time direction is expected
to be asymptotically flat and there should be a horizon present. The
coordinates $\phi$ and $\psi$ both have semi-infinite rod sources, so
these coordinates should be periodically identified and will have the
interpretation of azimuthal angles in the asymptotic metric. 

Consider the form of the 
metric as $y \rightarrow - \infty$. The $ty$ part of the metric
becomes
\be
 ds_{ty}^2 \sim F(x) \left( - \frac{1}{\mu |y|} dt^2 + \frac{\mu}{A^2
 |y|^3} dy^2 \right).
\ee
Performing the coordinate transformation
\be
 y = -\frac{4 \mu}{A^2 Y^2}
\ee
gives
\be
\label{eqn:nearhor}
 ds_{ty}^2 \sim F(x) \left( - \frac{A^2 Y^2}{4\mu^2} dt^2 + dY^2
 \right).
\ee
The metric in brackets is just that of Rindler space with
acceleration parameter $a=A/(2\mu)$. The coordinate transformation that
takes this to a manifestly flat metric is
\be
 X = Y \cosh at, \qquad T = Y \sinh at,
\ee
giving
\be
 ds_{ty}^2 \sim F(x)(-dT^2 + dX^2).
\ee
Note that the conformal factor $F(x)$ is always positive for $-1 \le x
\le 1$. This analysis shows that the leading order part of the $ty$
metric has a non-singular horizon at $y = -\infty$. If one examines
the subleading order terms, one finds that these are also regular
there if the same coordinate transformation is made.
It is easy to see that the other terms of the metric can also be
smoothly extended through this surface which is therefore a regular
horizon. The near-horizon metric is
\be
 ds^2 \sim F(x) \left( - dT^2 + dX^2 + A^{-2} d\psi^2 \right) +
 \frac{\mu^2}{A^2} \left( \frac{dx^2}{1-x^2} + \frac{1-x^2}{F(x)}
 d\phi^2 \right),
\ee
and the metric of a constant $t$ slice through the horizon is
\be
\label{eqn:horizonmetric}
ds^2 = \frac{1}{A^2} \left[ F(x) d\psi^2 + \mu^2 \left(
\frac{dx^2}{1-x^2} +
 \frac{1-x^2}{F(x)} d\phi^2 \right) \right].
\ee

Consider now the $x \phi$ part of the metric, which is conformal to
\be
 ds_{x\phi}^2 = \frac{dx^2}{1-x^2} + \frac{1-x^2}{F(x)} d\phi^2.
\ee
Let $x = -\cos \theta$ with $0 \le \theta \le \pi$. This
gives\footnote{
If $\mu = 1$ then the following analysis does not apply, but it is easy
to see that the $x \psi \phi$ part of the metric describes a round
$S^3$ of radius $2/A$ provided one identifies $\phi$ and $\psi$ with
period $2\sqrt{2} \pi$. This is consistent with the above comment that
the $\mu=1$ solution is just the five-dimensional Schwarzschild
solution.}
\be
 ds_{x\phi}^2 = d\theta^2 + \frac{\sin^2 \theta}{1 + \mu \cos \theta}
 d\phi^2.
\ee
In order for this metric to be regular at $\theta=0$ (i.e. $x=-1$), it
is necessary to identify $\phi$ with period $2\pi \sqrt{1+\mu}$. For
regularity at $\theta = \pi$ (i.e. $x=1$), it is necessary to identify
$\phi$ with period $2\pi \sqrt{1-\mu}$. It is therefore not possible
to have regularity at both $x=1$ and $x=-1$. If one demands regularity
at $x=-1$ then there is a conical singularity at $x=1$ with deficit
angle 
\be
 \delta_{(x=1)} = - 2\pi \left( \sqrt{\frac{1+\mu}{1-\mu}} - 1 \right),
\ee
which is negative so this is really an excess angle. If one demands
regularity at $x=1$ then there is a conical singularity at $x=-1$ with
deficit angle
\be
\label{eqn:delta2}
 \delta_{(x=-1)} = 2 \pi \left( 1- \sqrt{\frac{1-\mu}{1+\mu}} \right).
\ee
In both cases, the $ x \phi$ part of the metric describes a surface
that is topologically $S^2$ with a conical singularity at one of the
poles. In the full metric, this singularity is extended in two other
spatial dimensions and hence it describes a ``deficit membrane'', the
five-dimensional analogue of a four-dimensional deficit
string\footnote{
For a simpler example of a deficit membrane, consider the metric $ds^2
= -dt^2 + dr_1^2 + r_1^2 d\theta_1^2 + dr_2^2 + r_2^2 d\theta_2^2$
where $\theta_1$ is identified with period $2\pi$ and $\theta_2$ with
period $2\pi - \delta$. The deficit membrane sits at $r_2 = 0$.}. 

As $y \rightarrow -1$, $g_{\psi \psi}$ tends to zero. To analyze
this, set $y = - \cosh (\xi/\sqrt{1+\mu})$. Near $\xi = 0$, the
$y\psi$ part of the metric is conformal to
\be
 ds_{y \psi}^2 \approx d\xi^2 + \frac{\xi^2}{1+\mu} d\psi^2.
\ee
This is regular at $\xi = 0$ provided $\psi$ is identified with
period $\Delta \psi = 2 \pi \sqrt{1+\mu}$. $y=-1$ is then seen as the
origin of polar coordinates and hence $y$ cannot be continued beyond
$-1$. Returning to the horizon metric \ref{eqn:horizonmetric}, it is
now clear that the topology of the horizon is $S^1 \times S^2$, which
justifies calling this solution a black ring. The circumference of the
ring varies from a maximum of $2\pi (1+\mu) A^{-1}$ at $x=-1$ to a
minimum of $2 \pi \sqrt{1-\mu^2} A^{-1}$ at $x=1$. Since $x$ is the
polar coordinate on the $S^2$, it follows that $x=-1$ points away from
the ring and $x=+1$ points into the hole in the centre of the
ring. Thus the choice of where to put the conical deficit corresponds
either to having the black ring sitting on the rim of a disc shaped
deficit membrane (with negative deficit), or to it sitting on the rim of a
disc-shaped hole in an infinitely extended deficit membrane (with
positive deficit). The area of the horizon is
\be
 {\cal A}_h = 8 \pi^2 \frac{\mu^2 (1+\mu)}{A^{3}}
\ee
in the former case, and
\be
 {\cal A}_h = 8 \pi^2 \frac{\mu^2 \sqrt{1-\mu^2}}{A^{3}}
\ee
in the latter.

\begin{figure}
\begin{picture}(0,0)(0,0)
\put(170,160){$x$}
\put(362,284){$y$}
\put(240,280){$\psi$}
\put(50,145){$x=-1$}
\put(202,145){$x=1$}
\put(200,30){$y=-1$}
\put(330,170){$y=-\infty$}
\end{picture}
\centering{\psfig{file=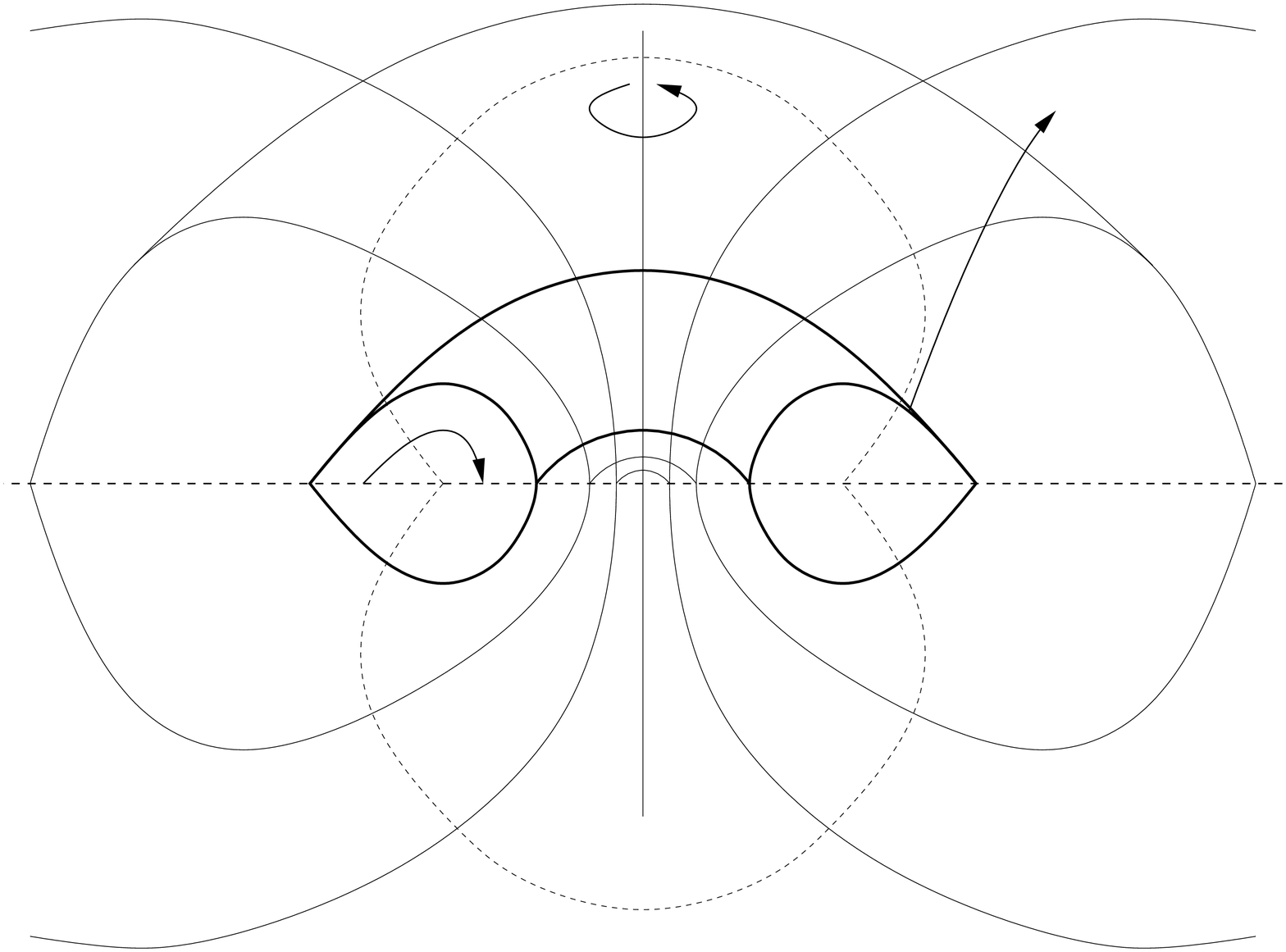,width=6.in}}
\caption{Spatial sections of the black ring metric. The
coordinate $\phi$ is suppressed. The surfaces of constant $y$ are
nested surfaces of topology $S^2 \times S^1$. The coordinate $\psi$ is
the coordinate on $S^1$. The coordinates $x$ and $\phi$ are,
respectively, the polar and azimuthal angles on $S^2$. The smallest
constant $y$ surface corresponds to the horizon,  at $y=-\infty$. The
surface at $y=-1$ degenerates into an axis of  rotation where the
orbits of $\psi$ shrink to zero. The surfaces of constant $x$ are
denoted by dotted lines. $x=-1$ points out of the ring and $x=+1$
points into the ring. The conical singularity may be chosen to lie
inside the ring or, as in the case shown, outside the ring (so that it
extends to infinity). Infinity is at $x=y=-1$.}
\label{fig:blackring}
\end{figure}

It is clear from the metric that the only values of $x$ and $y$ that
can correspond to asymptotic infinity are $x=y=-1$. As these values
are approached, the metric takes the asymptotic form
\be
 ds^2 \sim -dt^2 + \frac{1}{\tilde{A}^2 (x-y)^2} \left[ (y^2-1)
d\tilde\psi^2 + \frac{dy^2}{y^2-1} + \frac{dx^2}{1-x^2} + (1-x^2)
d\tilde\phi^2 \right],
\ee
where $\tilde\psi = \psi/\sqrt{1+\mu}$ and $\tilde\phi
=\phi/\sqrt{1+\mu}$ and $\tilde{A} = A/(1+\mu)$. The quantities
$\tilde\phi$, $\tilde\psi$, both have period $2\pi$ if the period of
$\phi$ is chosen such that the conical deficit lies at $x=1$. This
metric is in fact known to be flat space. The transformation
\be
 \xi = \frac{\sqrt{y^2-1}}{\tilde{A} (x-y)},\qquad
 \eta = \frac{\sqrt{1-x^2}}{\tilde{A} (x-y)},
\ee
takes it to the form
\be
 ds^2 \sim -dt^2 + d\xi^2 + d\eta^2 + \xi^2 d\tilde\psi^2 + \eta^2
 d\tilde\phi^2,
\ee
which is free of conical singularities if $\tilde\psi$ and $\tilde\phi$
both have period $2\pi$, which they do if the conical deficit lies at
$x=1$. If the conical deficit lies at $x=-1$ then $\tilde\phi$ has
a conical deficit $\delta_2$ given by equation \ref{eqn:delta2}, so
the asymptotic metric describes a flat deficit membrane in this case.
The structure of the black ring is summarized in figure
\ref{fig:blackring}. 

It has been shown that the black ring has an event horizon of topology
$S^2 \times S^1$. Naively, one might expect such a horizon to collapse
to form a spherical black hole horizon. However, the solution has
conical singularities that prevent this from occurring. These conical
singularities describe a deficit membrane that either extends to
infinity or forms a disc inside the ring. In the latter case, the
solution is asymptotically flat. We believe this to be the first
example of an asymptotically flat solution with an event horizon of
non-spherical topology. Of course, this solution requires the presence
of a conical {\it excess} angle, which corresponds to a deficit
membrane of negative tension. This is presumably unphysical, but it
will be shown in \cite{emparan:01} that the conical singularity can be
eliminated if the ring rotates in the $\psi$ direction.

If the asymptotic metric does not contain a conical singularity then
the mass of the black ring can be calculated by considering the
subleading contribution to $g_{tt}$. It is easy to show that near
$x=y=-1$ this behaves as
\be
 g_{tt} \sim - \left( 1 - \frac{2\mu (1+\mu)}{A^2 (\xi^2 +
\eta^2)} + \ldots \right),
\ee
from which it follows (see e.g. \cite{myers:86}) that the black ring has mass
\be
 M = \frac{3 \pi \mu (1+\mu)}{4 G_5 A^2},
\ee
where $G_5$ is Newton's constant in five dimensions. 
Note that when $\mu=1$ this gives the correct value for a
five-dimensional
Schwarzschild black hole of horizon radius $2/A$. 

If, on the other hand, the deficit membrane extends to infinity, the
mass of the ring can be calculated by taking as a reference background
the spacetime of a membrane (without a ring), with the result
\be
 M = \frac{3 \pi \mu \sqrt{1-\mu^2}}{4 G_5 A^2}.
\ee

The temperature of the black ring can be obtained by Euclideanizing the
near-horizon metric: $t = -i\tau$. In order to avoid a new conical 
singularity at $y = - \infty$, it is necessary to periodically
identify $\tau$. From equation \ref{eqn:nearhor}, one finds that
the temperature is
\be
 T = \frac{A}{4\pi \mu}.
\ee
The topology of the Euclidean solution is $(S^3 \times S^2) - S^1$, 
where the $S^3$ is covered by the coordinates $\tau$, $\psi$
and $y$, and the $S^2$ by $x$ and $\phi$. The circle removed is the
circle at $x=y=-1$ parametrized by $\tau$.

For either choice of the position of the deficit membrane, $x=1$ or
$x=-1$, there is a Smarr relation:
\be
\label{eqn:smarr}
 M = \frac{3}{8G_5}T{\cal A}_h.
\ee

Let us now assume the deficit membrane is outside the ring. The action
of the Euclidean solution can be computed similarly to
\cite{chamblin:97}, by subtracting the action of the deficit membrane
spacetime, and yields
\ba
I&=&{\pi^2\mu^2\sqrt{1-\mu^2}\over G_5 A^3}\nonumber\\
&=&{M\over 3 T}.
\ea
If we now identify the free energy as $F=TI=M-TS$, then using
\ref{eqn:smarr} we find that the
entropy satisfies the area law
\be
 S = \frac{{\cal A}_h}{4 G_5} = 
 \frac{2 \pi^2 \mu^2 \sqrt{1-\mu^2}}{G_5 A^3}.
\ee


It is unclear whether the black ring is a stable solution, or whether
it will become unstable for a certain range of parameter values. When
the radius of the $S^1$ grows to infinity we recover a translationally
invariant black string, which is known to be unstable
\cite{gregory:93}, and this suggests that the instability might set in
already for finite but large enough radius. In that case the ring would
be unstable to rippling along the $\psi$ direction. Given the presence
of the deficit membrane, a detailed analysis is needed to settle
the issue.

\subsection{Superposition of a black hole and KK bubble}

\label{sec:holebubble}

All of the metrics discussed above were already known\footnote{
Although the black ring metric had not been interpreted as such.} rather
than discovered using the general formalism of section
\ref{sec:weyl}. However, in this section and the following section,
new class II solutions will be constructed by following the steps 
described there. 

The first example is the $D=5$ solution that was labelled II(c) above.
It will be convenient to parametrize these sources slightly differently
from above, taking $U_1$ to be the potential of a finite rod $\mu/(2A)
\le z \le 1/(2A)$, $U_2$ to be the potential of a finite rod $-\mu/(2A)
\le z \le \mu/(2A)$ and $U_3$ to be the potential of the semi-infinite
rods $z \ge 1/(2A)$ and $z \le -\mu/(2A)$. The parameter $\mu$ will be
taken in the range $0 < \mu < 1$ (in order to prevent the rods from
overlapping). Using the same notation as
for the black ring, this gives
\be
 e^{2U_1} = e^{2u_1} \frac{R_2 - \zeta_2}{R_1 - \zeta_1},
\ee
\be
 e^{2U_2} = e^{2u_2} \frac{R_3 - \zeta_3}{R_2 - \zeta_2},
\ee
\be
 e^{2U_3} = e^{2u_3} (R_1 - \zeta_1) (R_3 + \zeta_3),
\ee
where the $u_i$ are arbitrary constants that reflect the freedom to
rescale the coordinates $x^i$. This will be used to avoid any conical
singularities along the axes: according to the general comments of
Section \ref{sec:interpret}, they can all be eliminated.

Following the prescription of section
\ref{sec:weyl}, one now computes the function $\nu$ by writing the
functions $U_i$ in terms of the complex coordinate $w = r+iz$ and then
integrating equations \ref{eqn:dwnu} and \ref{eqn:dwbarnu}. This
calculation is performed in appendix \ref{app:integrate}. The result
is
\be
 e^{2\nu} = \frac{e^{2\gamma_0}}{R_1 R_2 R_3} \sqrt{Y_{12} Y_{13}
 Y_{23}} \frac{R_1 - \zeta_1}{R_3 - \zeta_3},
\ee
where $\gamma_0$ is an arbitrary constant of integration. The quantities
$R_i$, $\zeta_i$ and $Y_{ij}$ are the same as for the black ring,
equations \ref{eqn:zidef}, \ref{eqn:Ridef}, \ref{eqn:Yijdef}. 

For the black ring,
the coordinate transformation $(r,z) \rightarrow (x,y)$ (defined by
equations \ref{eqn:rdef}, \ref{eqn:zdef} with $\alpha = A$) gives a form
of the metric that is easier to analyze. This suggests
performing the same coordinate transformation here.
The transformation of $dr^2 + dz^2$ under this change of coordinates can
be obtained from the above analysis of the black
ring. In these new coordinates, the metric takes the form 
\ba
 ds^2 &=& -\frac{F(x)}{F(y)} dt^2 + e^{2u_1} \frac{(1-x)F(y)}{(1-y)F(x)}
 (dx^1)^2 + \frac{e^{2u_3}}{A^4} \frac{(1+x) (1-y)^2 (-1-y) F(x)
 F(y)}{(x-y)^4} (dx^3)^2 \nonumber \\ {}&+& \frac{\sqrt{2} (1+\mu)
 e^{2\gamma_0}}{A^4 (x-y)^3} \left[ \frac{(1-y)F(x)}{-1-y} dy^2 +
 \frac{(1-y)^2 F(y)}{1-x^2} dx^2 \right],
\ea 
where the similarity with the black ring has suggested taking $x^2$ to
be the time direction, normalized such that $u_2 = 0$. The square of
the Riemann tensor diverges at $x = 1/\mu$, at $y=1/\mu$ and at
$y=1$. This suggests taking the ranges of the coordinates 
to be, again, $-1 \le x \le 1$ and $y \le -1$. 

This metric has an event horizon at $y = -\infty$,
just as for the black ring. The orbits of $x^1$ shrink to zero size at
$x=1$. Regularity requires that $x^1$ is
identified with period $2\pi$ and 
\be
 e^{2u_1} = 2\sqrt{2} (1-\mu^2) A^{-4} e^{2\gamma_0}.
\ee
The orbits of $x^3$ shrink to zero size at both $x=-1$ and $y=-1$. The
metric will be regular in both cases if $x^3$ is identified with
period $2\pi$ and
\be
 e^{2u_3} = 2\sqrt{2} e^{2\gamma_0}.
\ee
Having made these identifications, the above metric is complete and
non-singular and cannot be extended except through the event horizon at
$y=-\infty$. Note that $\gamma_0$ could be absorbed into $A$ and
can therefore be chosen to take any convenient value. The choice
\be
 e^{2\gamma_0} = \frac{A^2}{\sqrt{2}(1+\mu)}
\ee
will be made here. Letting $\phi = x^1$ and $\psi = x^3$, the metric
now takes the form
\ba
\label{eqn:holebubble}
 ds^2 &=& -\frac{F(x)}{F(y)} dt^2 + \frac{2(1-\mu)}{A^2}
 \frac{(1-x)F(y)}{(1-y)F(x)} d\phi^2 +
 \frac{2 (1+x) (1-y)^2 (-1-y) F(x) F(y)}{(1+\mu)A^2(x-y)^4} d\psi^2 
\nonumber \\
 {}&+& \frac{1}{A^2 (x-y)^3} \left[
 \frac{(1-y)F(x)}{-1-y} dy^2 + \frac{(1-y)^2 F(y)}{1-x^2} dx^2 \right],
\ea 
where $\phi$ and $\psi$ both have period $2\pi$. The event horizon at
$y=-\infty$ has topology $S^3$.

To interpret this metric, it is helpful to look first at certain
limiting cases. If $\mu \rightarrow 0$ then the sources for this
solution tend to the sources for a metric consisting of a flat time
direction times the $D=4$ Euclidean Schwarzschild solution. This is
the metric of a static KK bubble. If $\mu \rightarrow 1$ then the
sources tend to the sources for a $D=5$ black string, with $\phi$
becoming the translation coordinate along the string (one has to
rescale $\phi$ by $\sqrt{1-\mu}$ before taking $\mu \rightarrow
1$). Hence this metric must somehow interpolate between a static KK
bubble and a black string. 

Asymptotic infinity is at
$x=y=-1$. Near $x=y=-1$, the metric takes the form\footnote{
Near $x=y=-1$ one has $1-x \approx 2$ etc, however factors of $1-x$ etc
have been retained here for purposes of comparison with the KK bubble.}
\ba
 ds^2 &\sim & -dt^2 + \frac{2}{\tilde{A}^2} \frac{(1-x)}{(1-y)}
 d\tilde{\phi}^2 + \frac{2 (1+x) (1-y)^2 (-1-y)}{ \tilde{A}^2 (x-y)^4}
 d\psi^2 \nonumber \\ {} &+& \frac{1}{\tilde{A}^2 (x-y)^3} \left[
 \frac{(1-y)F(x)}{-1-y} dy^2 + \frac{(1-y)^2 F(y)}{1-x^2} dx^2 \right],
\ea 
where $\tilde{A} = A/\sqrt{1+\mu}$ and $\tilde{\phi} =
\sqrt{(1-\mu)/(1+\mu)} \phi$. This metric is what one would obtain
from the full metric \ref{eqn:holebubble}
with parameters $(\tilde{\mu},\tilde{A})$, where $\tilde{\mu}=0$. It
must therefore be the metric of a static KK bubble. The periodicity of
$\tilde{\phi}$ is inconsistent with regularity at $x=1$ but this metric
is only supposed to be an approximation to the metric
\ref{eqn:holebubble} near $x=y=-1$. The important point is that the
static KK bubble is known to be asymptotic to $R^{1,4} \times S^1$,
which is the KK vacuum metric. It follows that the metric
\ref{eqn:holebubble} must also be asymptotic to the KK vacuum, with the
KK circle parametrized by $\phi$. 

\begin{figure}
\begin{picture}(0,0)(0,0)
\put(75,225){\footnotesize\mbox{$\psi$}}
\put(75,205){\footnotesize\mbox{$x=-1$}}
\put(60,120){\footnotesize\mbox{$x=1$}}
\put(75,10){\footnotesize\mbox{$y=-1$}}
\put(50,110){\footnotesize\mbox{$y=-\infty$}}
\put(230,225){\footnotesize\mbox{$\psi$}}
\put(230,205){\footnotesize\mbox{$x=-1$}}
\put(215,100){\footnotesize\mbox{$x=1$}}
\put(230,10){\footnotesize\mbox{$y=-1$}}
\put(210,120){\footnotesize\mbox{$y=-\infty$}}
\end{picture}
\centering{\psfig{file=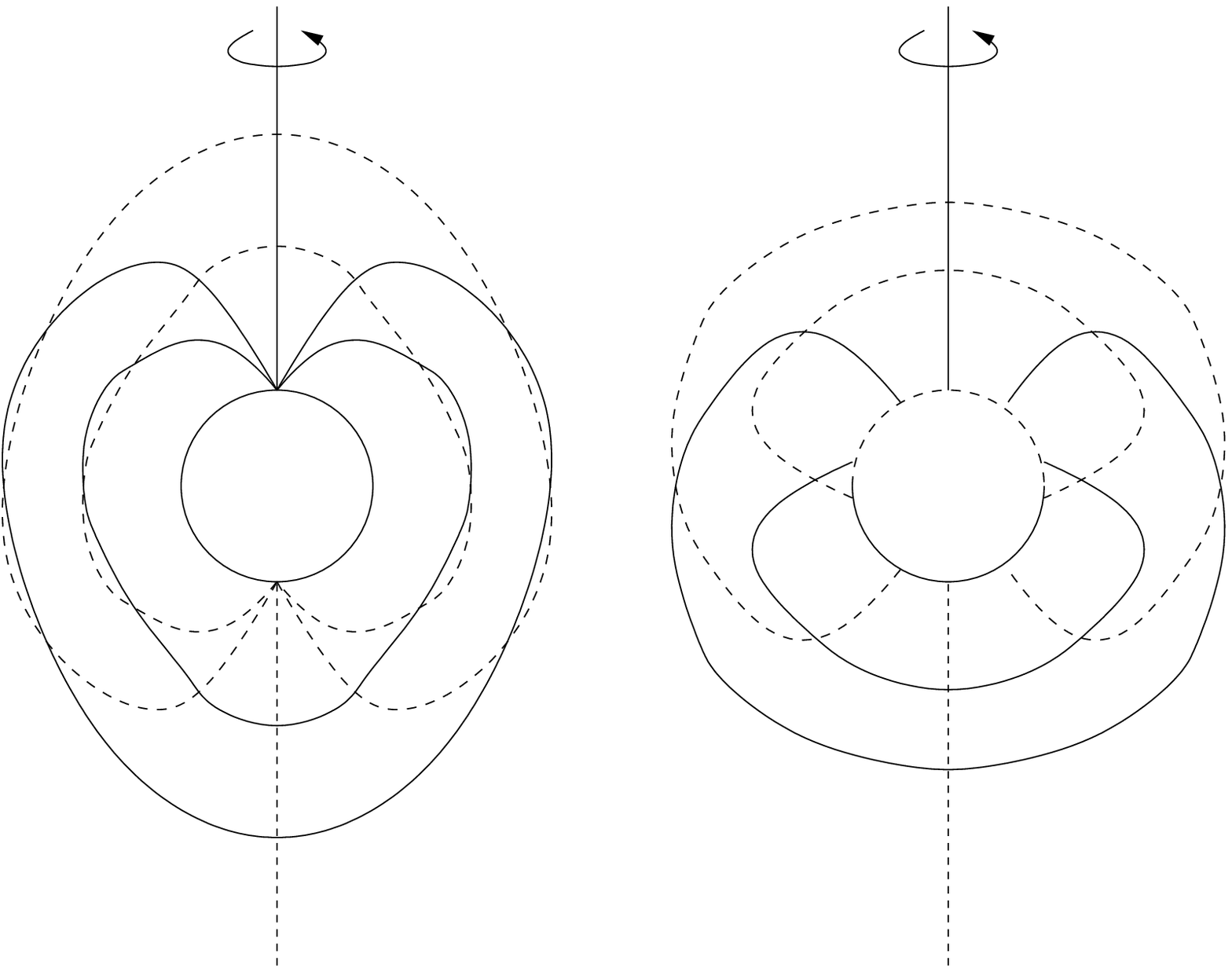,width=4.in}}
\caption{1. Schematic depiction of the $R\theta$ plane of the KK
bubble. The throat of the
bubble (where the KK circle shrinks to zero size) is at $R=2M$. 
In the $(x,y)$ coordinates, this corresponds to
$x=1$ or $y=-\infty$. The axes $\theta = 0,\pi$ correspond to $x=-1$
and $y=-1$ respectively. Solid and dashed lines denote curves of
constant $x$ and $y$ respectively. 2. The $xy$ plane of the metric
\ref{eqn:holebubble}. There is a horizon at $y=-\infty$ and the KK
circle shrinks to zero size at $x=1$.}
\label{fig:KKbubble}
\end{figure}

The orbits of $\psi$ shrink to zero size at $x=-1$ and at $y=-1$. To
understand what this means, it is convenient to consider the KK bubble
($\mu=0$) metric and how the coordinates $(x,y)$ relate to the
Schwarzschild coordinates $(R,\theta)$ in this case. This can be done
by setting $g_{\psi\psi}$ equal to\footnote{ The value of $M$ can be
fixed by looking at the sources in the Weyl form of the metric and
comparing with the Weyl form of the Schwarzschild metric: the lengths
of the rods should match.} $1-2M/R$ and $g_{\phi\phi}$
equal to $R^2 \sin^2 \theta$. Then  $x=-1$ corresponds to the axis
$\theta=0$ and $y=-1$ to the axis $\theta = \pi$. The surfaces of
constant $x$ and $y$ take the form shown on the left in figure
\ref{fig:KKbubble}. 

\begin{figure}
\begin{picture}(0,0)(0,0)
\put(-20,137){\footnotesize\mbox{$\phi=0$}}
\put(-20,91){\footnotesize\mbox{$\phi=\pi$}}
\put(140,95){\footnotesize\mbox{$x=1$}}
\put(70,140){\footnotesize\mbox{$x=-1$}}
\put(130,180){\footnotesize\mbox{$y=-1$}}
\end{picture}
\centering{\psfig{file=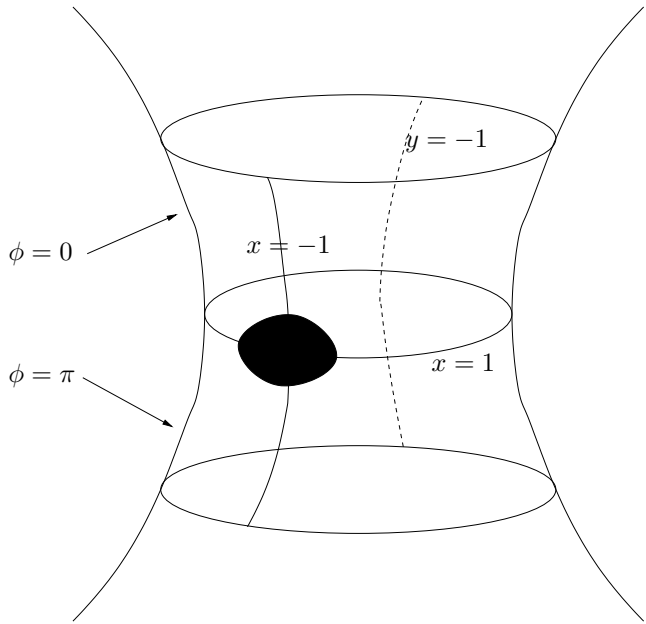,width=3.in}}
\caption{Geometry of the metric \ref{eqn:holebubble}. This picture shows
the surfaces $\phi=0$ (upper half) and $\phi=\pi$ (lower half), which
join together smoothly at $x=1$. The horizon at $y=-\infty$ corresponds
to a black hole sitting at the centre of the ``throat'' of a static
Kaluza-Klein bubble.}
\label{fig:throat}
\end{figure}

The metric \ref{eqn:holebubble} contains a horizon at $y=-\infty$. At
that point, the radius of the KK circle is finite.  This leads to
the picture on the right in figure \ref{fig:KKbubble}. 
The full geometry of the spatial sections can be visualized by
considering how the KK dimension varies. This is depicted in figure
\ref{fig:throat}. The solution describes a black hole sitting in the
``throat'' of a static KK bubble. 

In order to see that the topology of the horizon is $S^3$, note first
that the structure of the sources around the rod for the time
coordinate $x^2$ in figure \ref{fig:sources2}(c) is locally the same as
the rod structure in figure \ref{fig:sources1}(b) (with $x^1$ as time).
In more detail, note that for the KK bubble, the KK circle closes off
at $r=2M$ on a $S^2$. Let $\theta \in [0,\pi]$ denote the polar
coordinate on this sphere. If the black hole is now included then its
horizon intersects the $S^2$ at a circle (parametrized by $\psi$) at,
say, $\theta = \theta_*$ with the exterior region at $0 \le \theta <
\theta_*$. As one moves out of the throat, the $S^2$ expands, the KK
circle opens up and $\theta_*$ increases. Eventually, $\theta_*$
reaches $\pi$ and the horizon is no longer present. One can choose
coordinates on the horizon to be $\theta_*$, $\phi$ and $\psi$. At the
initial value of $\theta_*$, the circle parametrized by $\phi$ shrinks
to zero, and at the final value $\theta_*=\pi$, the circle parametrized
by $\psi$ shrinks to zero, from which it follows that the topology of
the horizon is $S^3$.
If $\mu \rightarrow 0$ then the black hole horizon shrinks to zero,
leaving a KK bubble. If $\mu \rightarrow 1$ then the horizon grows
until it swallows the throat of the bubble. When this happens, the KK
direction no longer closes off, and one is left with a compactified
black string with an event horizon of topology $S^2 \times
S^1$. 

If one Euclideanizes the solution then conical singularities can be
avoided if the Euclidean time direction $\tau$ is periodically
indentified with a period $\beta = 1/T$, corresponding to a
temperature
\be
 T = \frac{A}{4\pi\sqrt{\mu}}.
\ee
This instanton can probably be interpreted as describing an
instability of flat space at finite temperature in KK theory. 
This is a simultaneous manifestation of two different instabilities: the
bubble nucleation instability of the KK vacuum \cite{witten:82}
and the black hole nucleation instability of flat space at finite
temperature \cite{gross:82}. The
instanton might also be used to describe a decay of a compactified
black string by KK bubble nucleation. Presumably this instanton is not
allowed when fermions are included.

\begin{figure}
\begin{picture}(0,0)(0,0)
\end{picture}
\centering{\psfig{file=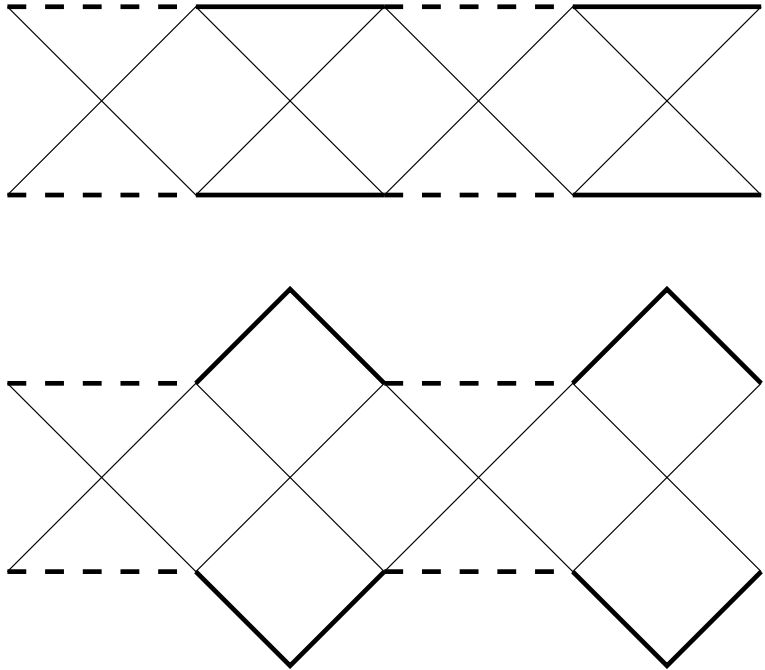,width=4.in}}
\caption{1. Causal structure of the two-dimensional metric
\ref{eqn:2dmetric}. The dotted lines denote curvature singularities,
the thick solid lines denote asymptotic infinity and the thin solid
lines denote horizons. The pattern can repeat indefinitely to the left
and right, or can be made finite by identifications.
2. Causal structure of the Wick rotated black
ring metric.}
\label{fig:penrose}
\end{figure}

It is known that the static KK bubble is classically unstable so it
seems likely that a similar instability will afflict this
solution. One might therefore wonder whether there is an analogue of
the expanding KK bubble solution (described by the Wick rotated $D=5$
Schwarzschild solution \cite{witten:82}) 
describing a black hole sitting in the throat
of the expanding bubble. Such a solution does indeed exist, and is
obtained by a Wick rotation of the black ring solution: if one lets
$t \rightarrow -i\tau$ and $\phi \rightarrow it$ in the black ring
solution then one obtains the metric\footnote{In terms of the original
notation for figure \ref{fig:sources2}(b), this means that the time
coordinate is $x^3$.}
\ba
\label{eqn:accbub}
 ds^2 &=& \frac{F(x)}{F(y)} d\tau^2 \\
 &+& \frac{1}{A^2 (x-y)^2} \left[ F(x)
 \left( (y^2-1) d\psi^2 + \frac{F(y)}{y^2-1} dy^2 \right) + F(y)^2
 \left( \frac{dx^2}{1-x^2} - \frac{1-x^2}{F(x)} dt^2 \right)
 \right]. \nonumber
\ea
The causal structure of this metric can be understood by first
examining the $xt$ part:
\be
\label{eqn:2dmetric}
 ds^2 = \frac{dx^2}{1-x^2} - \frac{1-x^2}{F(x)} dt^2.
\ee
By changing to Kruskal coordinates, it can be seen that there are
regular horizons at $x=\pm 1$ with different surface
gravities. The coordinates $(x,t)$ can be reintroduced beyond these
horizons. Continuing beyond the horizon at $x = 1$, one finds that
the square of the Riemann tensor diverges at $x = 1/\mu$. Beyond the
horizon at $x=-1$, the metric is asymptotically de Sitter.
Figure \ref{fig:penrose} shows the Carter-Penrose diagram for this
two-dimensional metric. The causal structure is the same as
Schwarzschild-de Sitter with the horizon at $x=1$ corresponding to the
black hole horizon and
the horizon at $x=-1$ corresponding to the cosmological horizon. As
for Schwarzschild-de Sitter it appears that there are many black holes
and asymptotic regions present, 
but one is free to identify these if one chooses. Doing so
clearly makes the spatial sections compact.

It is easy to see that the full five-dimensional metric will also have
regular horizons at $x = \pm 1$. The horizon at $x = 1$ has topology
$S^3$ and the one at $x = -1$ has topology $S^1 \times R^2$ (with the
$S^1$ direction parameterized by $\psi$). Continuing beyond the
horizon at $x=1$, there is a curvature singularity as
above. Asymptotic infinity is at $x=y$ and lies beyond the horizon at
$x=-1$. Near infinity, the metric approaches the KK vacuum with the KK
circle parametrized by $\tau$. The causal structure is illustrated in
figure \ref{fig:penrose}.\footnote{
Null infinity is presumably incomplete, as for the expanding KK bubble
\cite{dowker:95}.}
The interpretation of this metric is
that the horizon at $x=1$ is a black hole horizon (or horizons) and
the horizon at $x=-1$ an acceleration horizon that separates causal
curves that can fall into the black hole from those that cannot owing
to the expansion of space between them and the hole. This expansion of
space is just the expansion of the throat region of a KK bubble, which
is where the black hole is located. The asymptotic region beyond the
acceleration horizon is the region outside the bubble.

\subsection{Superposition of a black string and KK bubble}

\label{sec:stringbubble}

We parametrize the sources for the type II(d) solution as follows.
$U_1$ is the potential of a semi-infinite rod $z \ge 1/(2A)$, $U_2$ is
the potential of a semi-infinite rod $z \le -\mu/(2A)$, $U_3$ is the
potential of a finite rod $-\mu/(2A) \le z \le \mu/(2A)$ and $U_4$ is
the potential of a finite rod $\mu/(2A) \le z \le 1/(2A)$. This gives
\be
 e^{2U_1} = e^{2u_1} (R_1 - \zeta_1),
\ee
\be
 e^{2U_2} = e^{2u_2} (R_3 + \zeta_3),
\ee
\be
 e^{2U_3} = e^{2u_3} \frac{R_3 - \zeta_3}{R_2 - \zeta_2},
\ee
\be
 e^{2U_4} = e^{2u_4} \frac{R_2 - \zeta_2}{R_1 - \zeta_1},
\ee
where the $u_i$ are arbitrary constants. As in the previous
configuration, all conical singularities can be cancelled by an
appropriate choice of these constants and periodic identifications. 

The function $\nu$ is calculated using the method of appendix
\ref{app:integrate} with the result
\be
 e^{2\nu} = \frac{e^{2\gamma_0}}{R_1 R_2 R_3} \sqrt{Y_{12} Y_{23}}
 \sqrt{\frac{R_1 - \zeta_1}{R_3 - \zeta_3}},
\ee
where $\gamma_0$ is arbitrary. Once again, it proves useful to convert
from the Weyl coordinates $(r,z)$ to the black ring coordinates
$(x,y)$ using equations \ref{eqn:rdef} and \ref{eqn:zdef} (with
$\alpha = A$). This leads to
\ba
 ds^2 &=& -\frac{F(x)}{F(y)} dt^2 + e^{2u_4} \frac{(1-x)F(y)}{(1-y)F(x)}
 (dx^4)^2 \nonumber \\ &+& \frac{2 e^{2\gamma_0}}{A^2 (x-y)^2} \left\{ F(x)
 \left[\frac{dy^2}{-1-y} + \frac{1}{2} e^{2(u_1 - \gamma_0)} (y^2-1)
 (dx^1)^2 \right] \right. \\ 
 {} &+& \left. (1-y)F(y) \left[ \frac{dx^2}{1-x^2} + \frac{1}{2}
 e^{2(u_2 - \gamma_0)} (1+x) (dx^2)^2 \right] \right\}. \nonumber
\ea
The coordinate $x^3$ has been chosen as the time coordinate and
normalized so that $u_3 = 0$. The ranges of the coordinates will again
be taken to be $-1 \le x \le 1$ and $y \le -1$. 

This metric has an event horizon at $y=-\infty$. The orbits of
$x^2$ shrink to zero at $x=-1$. The metric will be regular there if
$x^2$ is indentified with period $2\pi$ and
\be
 e^{2u_2} = 4 e^{2\gamma_0}.
\ee
The orbits of $x^4$ shrink to zero at $x=1$. The metric will be
regular there if $x^4$ is identified with period $2\pi$ and
\be
 e^{2u_4} = 4(1-\mu) A^{-2} e^{2\gamma_0}.
\ee
The orbits of $x^1$ shrink to zero at $y=-1$. The metric will be
regular there if $x^1$ is identified with period $2\pi$ and
\be
 e^{2u_1} = 4 e^{2\gamma_0}.
\ee
The constant $\gamma_0$ could be absorbed into $A$ and can be
conveniently chosen as
\be
 e^{2\gamma_0} = \frac{1}{2}.
\ee
The metric therefore takes the form
\ba
\label{eqn:stringbubble}
 ds^2 &=& -\frac{F(x)}{F(y)} dt^2 + \frac{2(1-\mu)}{A^2} 
 \frac{(1-x)F(y)}{(1-y)F(x)}
 d\chi^2 \nonumber \\ &+& \frac{1}{A^2 (x-y)^2} \left\{ F(x)
 \left[\frac{dy^2}{-1-y} + 2(y^2-1)
 d\psi^2 \right] \right. \\ {} &+& \left. 
 (1-y)F(y) \left[ \frac{dx^2}{1-x^2} + 
 2(1+x) d\phi^2 \right] \right\}, \nonumber
\ea
where $\chi = x^4$, $\psi = x^1$ and $\phi = x^2$ all have period
$2\pi$. This metric is complete and non-singular outside of an event
horizon at $y=-\infty$ with topology $S^3 \times S^1$, where the $S^1$
is parametrized by $\psi$. 

This metric can be analyzed using arguments similar to those of the
previous section. It can be
seen that the metric is asymptotic to a $D=6$ static KK bubble
described by the product of a flat time direction with the $D=5$
Euclidean Schwarzschild solution. It follows that the metric is
asymptotic to the $D=6$ KK vacuum $M^{1,4} \times S^1$, with the $S^1$
parametrized by $\chi$. Note that the $S^1$ of the horizon does {\it not}
wrap the KK circle: spacelike infinity and the horizon both have
topology $S^3 \times S^1$ but in the former case, the $S^1$ is
parametrized by $\chi$ and in the latter by $\psi$.

Consider the spatial topology of the static KK bubble. The centre of the
bubble (where the KK direction collapses) has topology $S^3$. Moving
out of the bubble, the $S^3$ grows and the KK direction opens up so 
surfaces of constant radius from the bubble have topology 
$S^3 \times S^1_{\chi}$. For the above solution, this geometry is
altered by the presence of an event horizon. This event horizon
intersects the minimal $S^3$ of the static KK bubble on a $T^2$. To
see how this happens, introduce coordinates $(\theta,\phi,\psi)$ on
the $S^3$ such that $\phi$ and $\psi$ correspond to the coordinates used
above, $0 \le \theta \le \pi/2$ and the orbits of $\psi$ and $\phi$
collapse at $\theta=0$ and $\theta = \pi/2$ respectively. 
For example, the round metric on $S^3$ would take the form 
$ds^2 = d\theta^2 + \sin^2 \theta d\psi^2 + \cos^2 \theta d\phi^2$. 
The event horizon intersects the minimal $S^3$ at some value $\theta =
\theta_*$, so this intersection has topology $T^2 = S^1_{\psi} \times
S^1_{\phi}$. The metric outside the event horizon is at 
$0 \le \theta < \theta_*$. 

Moving away from the centre of the bubble, the $S^3$ expands, the KK 
circle $S^1_{\chi}$ opens up and $\theta_*$ increases. When $\theta_*$ 
reaches $\pi/2$, $S^1_{\phi}$ collapses to zero size. Beyond this
point, the event horizon no longer intersects the $S^3$. The event
horizon can therefore be parametrized by the coordinates $(\theta_*,
\chi, \psi, \phi)$. At the initial value of $\theta_*$, $S^1_{\chi}$
shrinks to a point and at the final value $\theta_* = \pi/2$,
$S^1_{\phi}$ shrinks to a point. $S^1_{\psi}$ remains finite over the
horizon. Hence the horizon has topology $S^3 \times S^1_{\psi}$, where
the $S^3$ is parametrized by $(\theta_*,\psi,\phi)$. 

In the limit $\mu \rightarrow 0$, the event horizon shrinks to nothing
and the metric reduces to the static KK bubble. As $\mu \rightarrow
1$, the event horizon grows to engulf the minimal $S^3$ and hence
there is nowhere that the KK direction collapses. In this limit, the
metric reduces to a black string wrapped around the KK direction, so
the event horizon has topology $S^3 \times S^1_{\chi}$.  

If the metric is Euclideanized by setting $t = -i \tau$ then conical
singularities can be avoided by identifying $\tau$ with period
$\beta$, corresponding to a temperature
\be
 T = \frac{A}{4 \pi \sqrt{\mu}}.
\ee
The topology of the Euclidean solution is $(S^3 \times S^3) - T^2$, 
where one $S^3$ is covered by the coordinates $(\tau,\psi,y)$
and the other by $(\chi,\phi,x)$ and the $T^2=S^1_{\chi} \times
S^1_{\phi}$ is at $x=y=-1$.

The static KK bubble is known to be unstable, which suggests that this
new solution is probably also unstable. For the black-hole bubble
solution discussed above, it was possible to obtain the solution
describing the evolution of the instability by Wick rotating the black
ring. In the present case, however, one can argue that such a solution,
if it exists, is not a class II solution, and perhaps not even a Weyl
solution.

\subsection{Other Wick rotations}

\label{sec:otherwick}

In order to complete the discussion of these new solutions, this section
will discuss the metrics obtained by Wick rotation. Thinking about the
sources for the Weyl solutions is useful in understanding what happens
when one Wick rotates the class II solutions. For example, in the
C-metric, $U_1$ and $U_2$ both have a finite rod and a semi-infinite rod
as sources. Therefore it does not matter whether $x^1$ or $x^2$ is
taken to be the time coordinate. For the black ring, the sources are
qualitatively different for each $U_i$ and hence the choice of which
$x^i$ is to be the time coordinate leads to physically distinct results.
These have all been discussed already. Taking $x^1$ to be the time
coordinate leads to the black ring. Taking $x^2$ to be the time
coordinate leads to the solution describing a pair of KK bubbles being
accelerated apart by a conical deficit \cite{chamblin:97}. Finally,
taking $x^3$ to be the time coordinate leads to the solution describing
an expanding KK bubble with a black hole sitting in the throat.

\begin{figure} \begin{picture}(0,0)(0,0)
\put(83,120){\footnotesize\mbox{$x=-1$}}
\put(-30,120){\footnotesize\mbox{$y=-1$}}
\put(10,35){\footnotesize\mbox{$x=1$}}
\put(50,35){\footnotesize\mbox{$y=-\infty$}} \put(220,5){\footnotesize
identify} \end{picture} \centering{\psfig{file=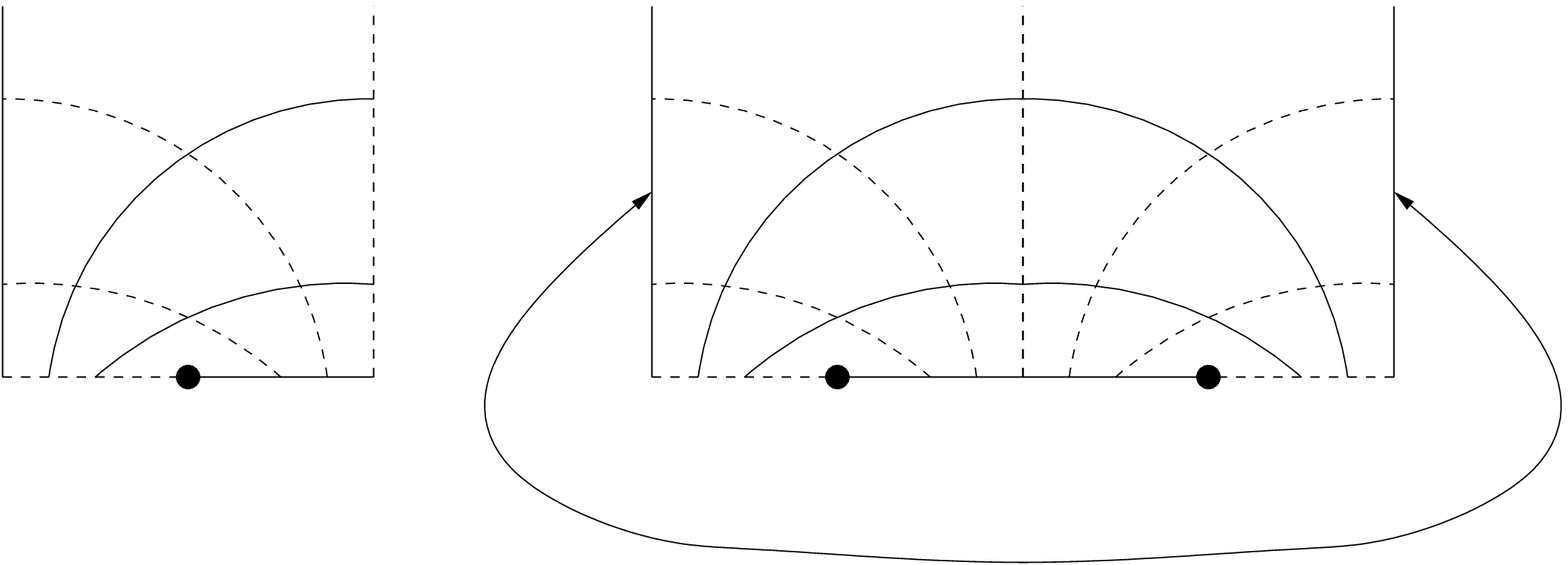,width=5.in}}
\caption{Spatial sections of the Wick rotated black hole-KK bubble
solution \ref{eqn:holebubble}. The KK coordinates $\phi$ and $\tau$ are
suppressed. The diagram on the left shows the region covered by the
coordinates $(x,y)$. The dotted and solid lines are curves of constant
$x$ and $y$ respectively. There are acceleration horizons at $x=-1$ and
$y=-1$. The $\phi$ direction closes off smoothly at $x=+1$ and the
$\tau$ circle closes off smoothly at $y=-\infty$. The heavy dots denote
points where both circles close off. The diagram on the right shows how
two copies of this region may be pasted together to give a complete
geometry.} \label{fig:wick1} 
\end{figure}

\begin{figure}
\begin{picture}(0,0)(0,0) \put(10,100){\footnotesize\mbox{$x=-1$}}
\put(260,100){\footnotesize\mbox{$y=-1$}}
\put(200,-5){\footnotesize\mbox{$x=1$}}
\put(70,-5){\footnotesize\mbox{$y=-\infty$}} \end{picture}
\centering{\psfig{file=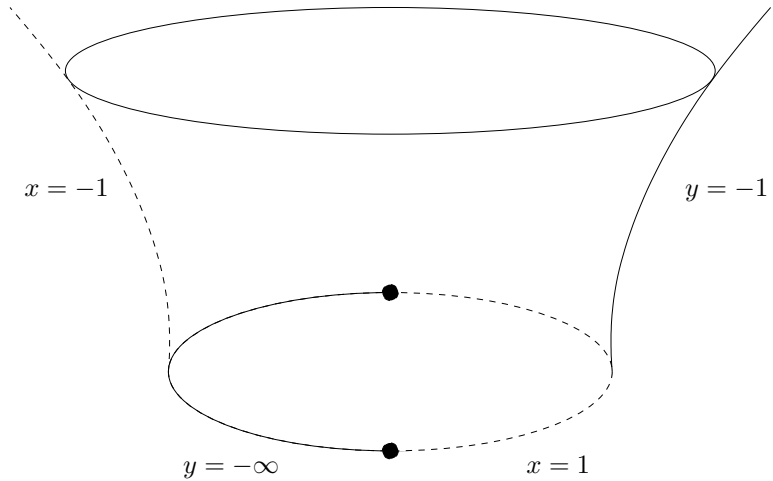,width=4.in}} \caption{Throat region of
the II(c) solution obtained by Wick rotation of the black hole$+$bubble
metric \ref{eqn:holebubble} to
$t\to -i\tau$, $\psi\to it$. The
$\phi$ and $\tau$ circles close off at $x=+1$ and $y=-\infty$
respectively, denoted by dotted and solid lines. Both circles close off
at the points denoted by heavy dots. As the throat expands, the distance
between these points increases and the acceleration horizons separate
regions which can receive light signals from each point.}
\label{fig:wick2} \end{figure}

For the new $D=5$ solution, II(c) (figure \ref{fig:sources2}(c)), it is
clear that taking $x^1$ or $x^2$ as the time coordinate gives physically
equivalent choices. Above we took $x^2$ as time, giving the static black
hole-KK bubble metric. However, taking $x^3$ to be the time direction
leads to a new metric. This is obtained from the
metric \ref{eqn:holebubble} by Wick rotating $t \rightarrow -i \tau$ and
$\psi \rightarrow i t$. The resulting metric is asymptotic to $M^{1,2}
\times T^2$ and has
acceleration horizons at $x=-1$ and $y=-1$. The KK directions are
parametrized by $\phi$ and $\tau$. Figure \ref{fig:wick1} shows the
geometry of the spatial sections with the KK directions suppressed.
The figure on the left shows
the region covered by the coordinates $(x,y)$. When the metric is
analytically continued beyond the acceleration horizon at $x=-1$ it
yields a new region isometric to the first. This can then be continued
beyond the horizon at $y=-1$ to yield yet a new region. Therefore, there
can be infinitely many such regions. By making identifications, the
number of regions can be made finite. For example, the figure on the
right shows how to identify in order to obtain just two regions. Figure
\ref{fig:wick2} shows the resulting spatial geometry. 
 The solution describes an expanding KK
bubble. The KK circles collapse to zero size in different regions of the
bubble's throat. These regions intersect in points. The acceleration
horizons separate the regions that these points can causally influence
as the throat expands. The Euclidean metric is an instanton for this
decay of the $M^{1,2} \times T^2$ KK vacuum. 

For the new $D=6$ solution, II(d) (figure \ref{fig:sources2}(d)) there
are only two physically inequivalent choices of the time coordinate. In
the discussion in Section \ref{sec:stringbubble} $x^3$ was chosen as the
time direction, giving the static black string-KK bubble metric. The
remaining possibility is to take, say, $x^2$ to be the time coordinate.
This corresponds to the Wick rotation $t \rightarrow -i \tau$, $\phi
\rightarrow it$ of the metric \ref{eqn:stringbubble}. The resulting
metric is asymptotic to $M^{1,3} \times T^2$, and has an acceleration
horizon at $x=-1$. The KK directions are parametrized by $\chi$ and
$\tau$, while $\psi$ is an azimuthal angle at infinity. One can
continue through the acceleration horizon as described above, leading
to an identical region. This solution can be interpreted as an
expanding KK bubble. The $\chi$ and $\tau$ circles collapse in
different regions of the bubble's throat. In a dimensionally reduced
picture, the KK bubble appears as a singularity of topology $S^2$,
with $\psi$ the azimuthal angle. 
The $\chi$ circle collapses near the poles of this sphere, and
the $\tau$ circle collapses on the rest of the sphere. The acceleration
horizon slices through the equator. If this metric is
Euclideanized then it gives an instanton for this decay of the
$M^{1,3} \times T^2$ vacuum.

\subsection{Different KK reductions}

\label{sec:newKK}

We have seen in section \ref{sec:interpret} that if a spatial
coordinate $x^i$ has only finite rod sources, then it is naturally
interpreted as a KK compactified direction. Above we have been
considering that points are identified along the orbits of the Killing
vector $\xi_{(i)}$. However, when there is more than one Killing
direction with compact orbits (whether their radius is asymptotically
constant or not) it is possible to perform the identifications along
the orbits of different linear combinations of the Killing vectors. Say
that $x^i$, $x^j$ are naturally identified with periodicities $\Delta
x^i$, $\Delta x^j$, in the sense that these identifications result in
the absence of (at least some) conical singularities along the axis of
symmetry. Then it would also be possible to, instead of identifying
points along the orbits of $\xi_{(i)}$, identify them along the orbits
of $\xi_{(i)}+{\Delta x^j\over\Delta x^i}\xi_{(j)}$, with no new
singularities arising. If the bolts of $\xi_{(i)}$ and $\xi_{(j)}$
intersect over a common fixed point, then, as shown in
\cite{dowker:96}, the circle action of this linear combination
generates a Hopf fibration of $S^3$. This change in the global
identifications obviously does not affect the local structure of the
solution, but it may result in a different interpretation of the
dimensionally reduced solution. An exhaustive study of this
construction has been performed in \cite{dowker:96}.

When  applied to the Weyl solutions, the most interesting case is that
where $x^i$ is a KK direction with asymptotically constant radius, and
$x^j$ is an azimuthal angle. In other words, 
$U_i$ has only finite rod sources,
and $U_j$ has at least one semi-infinite rod source. In this case, the
twisted KK circle action is interpreted as the Hopf fibration of a
magnetic monopole. The isolated fixed point of the fibration ---the
common fixed point of $\xi_{(i)}$ and $\xi_{(j)}$--- appears, in the
reduced spacetime, as the (singular) source of a magnetic field.

To illustrate this with an example \cite{dowker:96}, consider adding a
flat time direction to the $D=4$ Euclidean Schwarzschild solution
(refer to section \ref{sec:schwarz} and figure \ref{fig:sources1}(a)).
$x^1$ is a KK direction, with natural periodicity $8\pi M$, and $x^2$
is an azimuthal angle of period $2\pi$. With the conventional
(untwisted) identifications  this describes a static KK bubble.
However, $\xi_{(1)}+(1/4M)\xi_{(2)}$ generates Hopf actions with
opposite orientations around the endpoints of the rod. Identifying
points along these orbits, these endpoints appear in the reduced
four-dimensional description as a pair of oppositely charged magnetic
monopoles. This reduced spacetime is not asymptotically flat: the
change in the identifications results also in a KK magnetic Melvin
flux tube \cite{dowker:94}, which balances the attraction between
the monopole and the antimonopole. Since the strength of the external
magnetic flux tube and the charge of the monopoles are both determined
by the amount of twist in the reduction, they are not independent
parameters. 

Proceeding this way we are led to alternative interpretations of many
of the Weyl solutions that contain KK bubbles. The $D=5$ Schwarzschild
solution was found to describe in this manner a pair of oppositely
charged magnetic monopoles accelerating away under the pull of a
magnetic flux tube \cite{dowker:95, dowker:96}. Consider now the
solutions in class II(b) (refer to figure \ref{fig:sources2}(b)), with
$x^1$ as the KK coordinate. Dimensionally reduce along the orbits of
$\xi_{(1)}+{\Delta x^3\over\Delta x^1}\xi_{(3)}$, so $x^2$ is the
timelike coordinate. The endpoints of the $U_1$ rod have the
four-dimensional interpretation of a magnetic monopole and antimonopole.
$U_2$ has a semi-infinite rod, so $x^2$ is a boost coordinate and the
pair are accelerating, but notice they do it {\it together}. By
extending the solution across the acceleration horizon, we expect to
find a similar monopole-antimonopole dipole accelerating in the opposite
direction. The dipoles accelerate under the pull of a magnetic
flux tube, and each dipole is held together by the presence of a conical
singularity running between the poles.

Now suppose we reduce along the orbits of $\xi_{(1)}+{\Delta
x^2\over\Delta x^1}\xi_{(2)}$. From the sources for the time coordinate
$x^3$ we infer the presence of an acceleration horizon, but also of a
black hole horizon. The vectors $\xi_{(1)}$ and $\xi_{(2)}$ do not
share common fixed points: in this case the magnetic charge is not
sourced by a monopole, but by a black hole. The interpretation is in
terms of a non-extremal KK magnetic black hole moving with uniform
acceleration (and its opposite counterpart, from analytic continuation,
beyond the acceleration horizon). Since, as we explained, the charge
and the magnetic field cannot be varied independently of each other,
the solution is a only particular case of the magnetic KK Ernst
solution of \cite{dowker:94}.

In a similar vein, the II(c) solution with identifications along
$\xi_{(2)}+{\Delta x^3\over\Delta x^2}\xi_{(3)}$ leads to a static
configuration with a magnetic monopole and an oppositely charged
magnetic black hole. A flux tube keeps them apart in (unstable)
equilibrium. The II(d) solution, with an interpretation in terms of
$T^2$ compactified KK theory, admits even more combinations, which we
shall leave to the reader to analyze.

\subsection{Multi-black hole configurations}

\label{sec:multibh}

Weyl's construction in $D=4$ easily allows for configurations with an
arbitrary number of black holes along the symmetry axis: these are the
Israel-Khan solutions \cite{israel:64}. As is obvious from physical
considerations, the conical singularities can only be cancelled if there
are an infinite number of them (and then the masses and distances
between them are properly adjusted). With such a periodic array of black
holes, it is natural to periodically identify the $z$-coordinate,
which gives a solution describing a black hole localized on the KK
circle of $D=4$ KK theory \cite{myers:87}. 

It is a simple matter to describe Weyl solutions in $D>4$ with
several disconnected horizons. This provides the first example of a
construction of static multi-black hole solutions in higher dimensional
vacuum gravity. Here we will only sketch the properties of the
solutions as deduced from their rod structure, rather than giving the 
full metrics.

\begin{figure}
\centering{\psfig{file=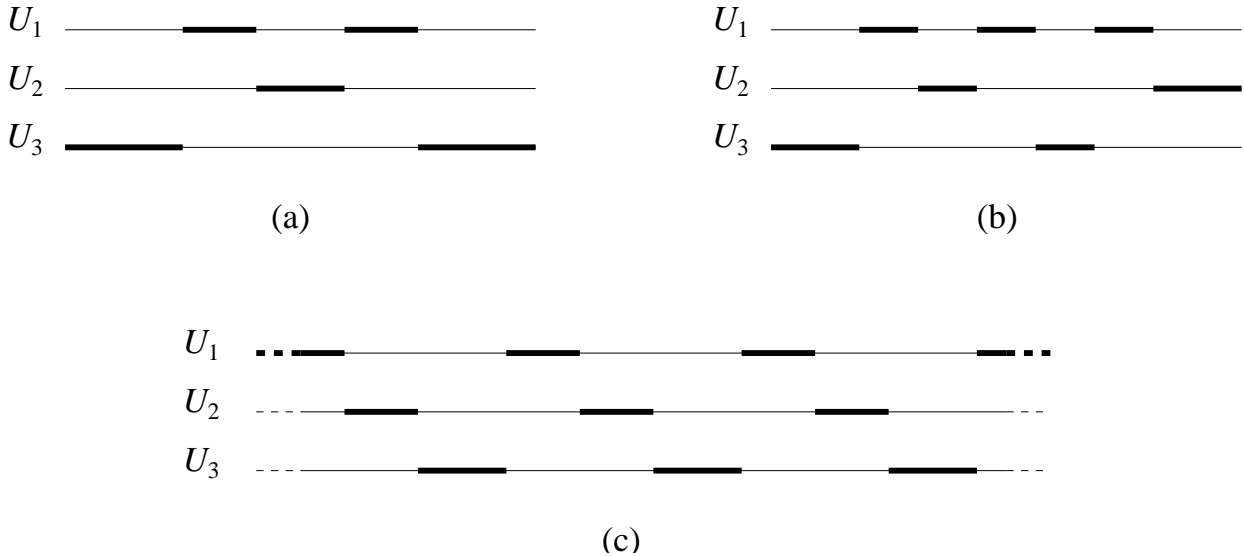,width=6.5in}}
\caption{Sources for (a) two-black hole configuration; (b) three-black
hole configuration; (c) infinite periodic array of black holes.}
\label{fig:sourcesn}
\end{figure}

The simplest solution with two black hole horizons, each of topology
$S^3$, results from the class III sources shown in figure
\ref{fig:sourcesn}(a). However, due to the isolated finite rod source
for $U_2$, the coordinate $x^2$ is asymptotically a KK circle, so this
is not an asymptotically flat solution. Instead, the configuration
describes two black holes at the north and south pole of a KK bubble:
add a second black hole to figure \ref{fig:throat}, sitting opposite to
the one that is already present. The distribution of the sources also
reveals that no conical  singularities are needed to keep the black
holes apart. When the two black holes coalesce, the solutions does not
reduce to  a single, larger, black hole but rather to a black string.

A three-black hole solution is obtained from the sources of figure
\ref{fig:sourcesn}(b). This solution {\it is} asymptotically flat, with
the two spatial Killing directions $x^2$, $x^3$ becoming azimuthal
angles at infinity. It necessarily contains conical singularities.
However, the black holes cannot be described as collinear. The first and
second black holes (numbering sources from the left) lie at the north
and south pole of a topological $S^2$ parameterized by $z$ and $x^3$,
while the second and third black holes lie similarly on another
topological $S^2$ parameterized by $z$ and $x^2$. So the second black
hole is collinear with each of the other two, but along different axes.
If two of the black holes coalesce, then we find a configuration of a
black hole encircled by a black ring.

Solutions with an infinite number of black hole horizons can be
obtained in several ways. The simplest possibility is shown in figure
\ref{fig:sourcesn}(c). It is likely that conical singularities can be
eliminated from this solution if each $U_i$ has sources consisting of
rods of equal length. It is natural to periodically identify the
$z$-axis to obtain a solution with a single black hole localized
on a KK circle, parametrized by $z$, of fixed length at infinity.
The coordinates $x^2$ and $x^3 $ also parametrize circles but these do
not approach a constant length at infinity so they cannot be regarded
as KK directions. Spacelike infinity has topology $S^1_2 \times S^1_3
\times S^1_z$ rather than $S^2 \times S^1_z$ which would be
appropriate for a description of a black hole localized on a KK
circle. It is clear that other similar configurations of infinitely many rods
will suffer from the same drawback.

Configurations with multiple concentric black rings are also possible,
but we shall stop at this point.

\sect{Concluding remarks}

\label{sec:discussion}

We have succeeded in generalizing Weyl's class of
solutions to arbitrary dimension
by finding the general solution of the vacuum Einstein
equations in $D$-dimensions that admits $D-2$ orthogonal commuting
non-null Killing vector fields. 
There are two classes of static Weyl solutions. The first (the ``generic''
Weyl solutions) is parametrized by
$D-3$ independent axisymmetric harmonic functions in three dimensional
flat space. The second class (the ``special'' Weyl solutions), 
analyzed in Appendix \ref{app:special},
is parametrized by $D-4$ independent harmonic functions in two
dimensional flat space. All known physically relevant solutions fall
into the first class, and these solutions were all found to have
harmonic functions produced by sources consisting of thin rods on the
$z$-axis, with mass $1/2$ per unit length. A natural way of
classifying the solutions was proposed, according to the number of
finite rod sources for the harmonic functions. This classification
scheme presented three new solutions as promising candidates for
deeper study.

Perhaps the most interesting new solution is the black ring of Section
\ref{sec:ring}. This is the first example of a stationary 
solution of the vacuum
Einstein equations that is asymptotically flat and has an event horizon
of non-spherical topology\footnote{A toroidal horizon in
four dimensions has been observed in numerical simulations of collapse
in \cite{hughes:94}. It is a transient phase of the collapse: the hole
in the torus closes up faster than the speed of light, thereby
preventing asymptotic observers from probing the topology of the horizon
\cite{shapiro:95}.}. The black ring is supported against collapse by a
conical deficit singularity in the form of a disc that sits inside the
ring. This singularity might be regarded as the gravitational effect of
a thin membrane of matter, much as a deficit string can be regarded as
an idealization of a real cosmic string. The deficit for the black ring
has {\it negative} deficit angle, and hence corresponds to a negative
tension source. It is not likely that it can be modelled by any
reasonable matter source, since the weak energy condition would not be
obeyed. An alternative which is perhaps physically more reasonable is to
take the conical deficit to lie outside the ring. A deficit membrane of
positive tension is then present which extends to infinity, so the
solution is no longer asymptotically flat. 

If the black ring were charged, one might envisage balancing it against
collapse (and therefore cancelling the conical singularity) by immersing
it in a background field. This is actually the situation with the
five-dimensional charged black ring solution of \cite{emparan:01a},
which is the first example of a stationary solution with spacelike infinity of
spherical topology and a regular horizon of non-spherical topology. The
horizon of the ring in that case is an extremal, degenerate one, with
vanishing horizon area. It is nevertheless completely non-singular. The
presence of the background field implies that the spacetime is not
asymptotically flat, instead it asymptotes to a fluxbrane solution. An
alternative to coupling to a background field is to set the ring into 
rotation. It will be shown in \cite{emparan:01} that one can indeed
obtain a vacuum solution describing an asymptotically flat rotating
black ring\footnote{Other recent solutions where rotation plays a role
in balancing charged, ring-like or tube-like configurations, have been
given in \cite{lunin:01}.} that is free of conical singularities.

The other new solutions contain KK bubbles in addition to black holes
or black strings. Some of them describe new decays of the KK vacua with
internal $S^1$ or $T^2$. We have exhibited but a few examples of the
wide range of possibilities for solutions which are singular in the KK
reduced description but nevertheless completely regular in higher
dimensions. Just as the black ring arose from a reinterpretation of a
previously known solution, maybe the other class II (or higher)
solutions have unexpected applications.

We have sketched the construction of multi-black hole configurations. In
contrast to the Israel-Khan solutions in $D=4$, the generalized Weyl
solutions cannot describe a linear array of five-dimensional black holes
(which would have symmetry ${\bf R}\times O(3)$ instead of ${\bf
R}\times O(2)^2$). Nevertheless, we have given the first examples of
static vacuum multi-black hole configurations in dimension higher than
four. Unlike the Israel-Khan solutions, some of the higher dimensional
solutions with finitely many black holes do not contain conical
singularities.

\begin{table}[tb]
\begin{center}
\caption{Summary of the main solutions studied in this paper, in the
classification of sec.~\ref{sec:classes}. The sources and metrics are
referred to the figures and equations in this paper, respectively. Only
the topology of finite area horizons is described. Other
interpretations, by different Wick rotations
or KK reductions, are discussed in secs.~\ref{sec:otherwick} and
\ref{sec:newKK}.}
\medskip
\begin{tabular}{ccclcc}
\hline
Class  & Sources (Fig.) & Metric (Eq.) &  Interpretation 
& Horizon	\\
\hline \hline
0	& 1	&\ref{eqn:flat} & Flat (Rindler) space  & - \\ 
I	& 2(a) 	& -	 	& 4D Black hole		& $S^2$ \\
{}	& 2(b) 	&\ref{eqn:5dsch}& 5D Black hole		& $S^3$	\\
II	& 3(a) 	& -		& C-metric		& $S^2$\\
{}	& 3(b) 	&\ref{eqn:bring}& Black ring 		
& $S^2\times S^1$\\
{}	& 3(b) 	&\ref{eqn:accbub}& Black hole in expanding KK bubble 
& $S^3$\\
{}	& 3(c) 	&\ref{eqn:holebubble}& Black hole in static KK bubble 
&$S^3$ \\
{}	& 3(d) 	&\ref{eqn:stringbubble}& Black string in static KK
bubble  & $S^3\times S^1$\\
$>$II	& 10(a,b,c,d)& - & Multi-black holes &	
$\bigoplus_{n} S^3$ \\
\hline
\end{tabular}
\end{center}
\label{table}
\end{table}

There are several directions for extensions of the Weyl classes in this
paper. Consideration of non-orthogonal Killing vectors would lead to
non-static, stationary solutions, or solutions with twists among the
axes. However, in $D=4$ there is no general solution for non-orthogonal
Killing vectors, so progress here could probably only be made in special
cases. It would also be interesting to study the addition of $p$-form
gauge fields.

To conclude, we have performed a systematic analysis of an infinite
class of exact solutions of vacuum gravity in dimensions higher than
four, and exhibited new solutions with qualitatively new properties. 
We hope that this work helps stimulate further systematic studies on the
rich structure of exact solutions of higher-dimensional gravity.

\bigskip

\centerline{\bf Acknowledgments}

We are grateful to Andrew Chamblin, Fay Dowker, Jerome Gauntlett and
Gary Gibbons for discussions. RE
acknowledges partial support from UPV grant 063.310-EB187/98 and CICYT
AEN99-0315. HSR was supported by PPARC.

\appendix

\setcounter{equation}{0}
\renewcommand{\theequation}{\Alph{section}.\arabic{equation}}

\sect{Curvature components}

\label{app:curvature}

Introduce a vielbein for the metric of equation \ref{eqn:lineelement}:
\be
 e^i = e^{U_i} dx^i, \qquad e^Z = e^C dZ, \quad e^{\bar{Z}} = e^C
 d\bar{Z}.
\ee
The summation convention is not being used for the indices $i.j.\ldots$.
The tangent space metric $\eta_{\alpha \beta}$ is given by $\eta_{ii}
= \epsilon_i$, $\eta_{Z \bar{Z}} = \eta_{\bar{Z} Z} = 1/2$, with other
components vanishing. The connection 1-forms are defined by
\be
 de^{\alpha} = - \omega^{\alpha}{}_{\beta} \wedge e^{\beta},
\ee
and explicit calculation gives
\be
 \omega_{iZ} = e^{-C} \partial_Z U_i e_i, \qquad \omega_{i\bar{Z}} =
 e^{-C} \partial_{\bar{Z}} U_i e_i, \qquad \omega_{ij} = 0,
\ee
\be
 \omega_{Z\bar{Z}} = -\frac{1}{2} e^{-C} \partial_Z C e^Z +
\frac{1}{2} e^{-C} \partial_{\bar{Z}} C e^{\bar{Z}}.
\ee
The curvature 2-forms are defined by
\be
 \Theta_{\alpha\beta} = d\omega_{\alpha \beta} + \omega_{\alpha}
 {}^{\gamma} \wedge \omega_{\gamma \beta}.
\ee
The non-vanishing curvature 2-forms are
\be
 \Theta_{ij} = -2 e^{-2C} \left( \partial_Z U_i \partial_{\bar{Z}} U_j
 + \partial_Z U_j \partial_{\bar{Z}} U_i \right) e_i \wedge e_j,
\ee
\ba
\Theta_{iZ} &=& -e^{-2C} \left[\partial^2_Z U_i + (\partial_Z U_i)^2 -
 2 \partial_Z C \partial_Z U_i \right] e_i \wedge e^Z \nonumber \\
 &-& e^{-2C} \left[\partial_Z \partial_{\bar{Z}} U_i + \partial_Z U_i
 \partial_{\bar{Z}} U_i \right] e_i \wedge e^{\bar{Z}},
\ea
\ba
 \Theta_{i\bar{Z}} &=& - e^{-2C} \left[\partial_Z \partial_{\bar{Z}}
 U_i + \partial_Z U_i \partial_{\bar{Z}} U_i \right] e_i \wedge e^Z
 \nonumber \\
 &-& e^{-2C} \left[ \partial^2_{\bar{Z}} U_i + (\partial_{\bar{Z}}
 U_i)^2 - 2\partial_{\bar{Z}} C \partial_{\bar{Z}} U_i \right] e_i
 \wedge e^{\bar{Z}},
\ea
\be
\Theta_{Z \bar{Z}} = e^{-2C} \partial_Z \partial_{\bar{Z}} C e^Z \wedge
 e^{\bar{Z}}. 
\ee
The tangent space components of the Riemann tensor are obtained from these
expressions by
\be
 \Theta_{\alpha \beta} = \frac{1}{2} R_{\alpha \beta \gamma \delta}
 e^{\gamma} \wedge e^{\delta},
\ee
with the results
\be
 R_{ijkl} = -2e^{-2C} \left( \partial_Z U_i \partial_{\bar{Z}} U_j +
 \partial_Z U_j \partial_{\bar{Z}} U_i \right) \left( \eta_{ik}
 \eta_{jl} - \eta_{il} \eta_{jk} \right),
\ee
\be
 R_{iZjZ} = -e^{-2C} \left[ \partial^2_Z U_i + (\partial_Z U_i)^2 -
 2 \partial_Z C \partial_Z U_i \right] \eta_{ij},
\ee
\be
 R_{i\bar{Z} j \bar{Z}} = - e^{-2C} \left[ \partial^2_{\bar{Z}} U_i +
 (\partial_{\bar{Z}} U_i)^2 - 2\partial_{\bar{Z}} C \partial_{\bar{Z}} 
 U_i \right] \eta_{ij},
\ee
\be
 R_{iZ j \bar{Z}} = -e^{-2C} \left( \partial_Z \partial_{\bar{Z}} U_i
 + \partial_Z U_i \partial_{\bar{Z}} U_i \right) \eta_{ij},
\ee
\be
 R_{Z\bar{Z} Z \bar{Z}} = e^{-2C} \partial_Z \partial_{\bar{Z}} C,
\ee
with any other non-vanishing components related to these by the
symmetries of the Riemann tensor. The non-vanishing tangent space
components of the Ricci tensor are given by
\be
 R_{ij} = -2 e^{-2C} \left[ 2 \partial_Z \partial_{\bar{Z}} U_i +
 \partial_Z U_i \sum_k \partial_{\bar{Z}} U_k + \partial_{\bar{Z}} U_i \sum_k
 \partial_Z U_k \right] \eta_{ij},
\ee
\be
 R_{ZZ} = -e^{-2C} \sum_i \left( \partial^2_Z U_i + (\partial_Z U_i)^2 -
 2 \partial_Z C \partial_Z U_i \right),
\ee
\be
 R_{\bar{Z} \bar{Z}} = -e^{-2C} \sum_i \left(\partial^2_{\bar{Z}} U_i +
 (\partial_{\bar{Z}} U_i)^2 - 2\partial_{\bar{Z}} C \partial_{\bar{Z}} 
 U_i \right),
\ee
\be
 R_{Z \bar{Z}} = -e^{-2C} \left[ 2 \partial_Z \partial_{\bar{Z}} C +
 \sum_i \partial_Z \partial_{\bar{Z}} U_i + \sum_i \partial_Z U_i
 \partial_{\bar{Z}} U_i \right].
\ee

\sect{The special Weyl solutions}
\label{app:special}

When solving the Einstein equations in Section \ref{sec:solving}, a
specific assumption was made in order to deal with the $ZZ$ and
$\bar{Z}\bar{Z}$ components: that the functions $w(Z)$ and
$\tilde{w}(\bar{Z})$ that appear in \ref{eqn:wdef} are non-constant. In
this appendix we investigate the cases in which one or both of these
quantities is constant. Consider first the case in which $Z$ and
$\bar{Z}$ are complex conjugate coordinates. Then\footnote{ Note that
$w(Z)^* \equiv \bar{w}(\bar{Z})$.} $\tilde{w}(\bar{Z}) = w(Z)^*$ so both
$w$ and $\tilde{w}$ must be constant. The $ij$ components of the vacuum
Einstein equations therefore reduce to
\be
 \sum_i U_i = {\rm constant},
\ee
\be
 \partial_Z \partial_{\bar{Z}} U_i = 0.
\ee
These equations have the solution
\be
 U_i(Z,\bar{Z}) = a_i(Z) + a_i(Z)^*,
\ee
with $a_i(Z)$ arbitrary except for the constraint
\be
\label{eqn:aicons1}
 \sum_i a_i(Z) = {\rm constant}.
\ee
The $ZZ$ and $\bar{Z}\bar{Z}$ components of the Einstein equation
reduce to 
\be
\label{eqn:aicons2}
 \sum_i (\partial_Z a_i)^2 = 0.
\ee
The $Z\bar{Z}$ component of the Einstein equation can be written
\be
 \partial_Z \partial_{\bar{Z}} \left(C + \frac{1}{4} \sum_i U_i^2
\right) = 0,
\ee
with solution
\be
 C(Z,\bar{Z}) = -\frac{1}{4} \sum_i U_i^2 + c(Z) + c(Z)^*,
\ee
where $c(Z)$ is arbitrary. This arbitrary function just reflects the
freedom to change coordinates $Z \rightarrow Z'(Z)$. Thus the distinct
solutions are labelled by the functions $a_i(Z)$. These $D-2$ functions
are constrained by equations \ref{eqn:aicons1} and \ref{eqn:aicons2},
so only $D-4$ of them are independent. For $D=4$ the solution is flat
space. For $D>4$ the solutions are non-trivial and have no
four-dimensional analogue. Note that each $U_i$ is a solution of the
Laplace equation in two-dimensional flat space, so this special class
of solutions is determined by $D-4$ harmonic functions in two flat
dimensions, in contrast to the class of generic Weyl solutions
discussed in Section \ref{sec:solving}, which was determined in terms
of $D-3$ axisymmetric harmonic functions in three flat dimensions.
In $D=5$, it is possible to explicitly solve the constraints on the
$a_i$'s to obtain a line element parametrized by an arbitrary function
of one complex coordinate.

Now consider the case in which $Z$ and $\bar{Z}$ are independent real
coordinates and $\tilde{w}$ is constant but $w(Z)$ is not. $\tilde{w}$
can be absorbed into $w(Z)$. The $ij$ Einstein equations reduce to
\be
 \sum_i U_i = \log (w(Z)),
\ee
\be
 2 w \partial_Z \partial_{\bar{Z}} U_i + \partial_Z w
\partial_{\bar{Z}} U_i = 0,
\ee
with solution
\be
 U_i(Z,\bar{Z}) = a_i(Z) + \tilde{a}_i (\bar{Z}) w^{-1/2},
\ee
where $a_i(Z)$ and $\tilde{a}_i(\bar{Z})$ are arbitrary except for the
constraints
\be
\label{eqn:aitcons1}
 \sum_i \tilde{a}_i(\bar{Z}) = \tilde{a},
\ee
\be
\label{eqn:aicons3}
 \sum_i a_i(Z) = \log w - \tilde{a} w^{-1/2},
\ee
where $\tilde{a}$ is a constant. The $\bar{Z}\bar{Z}$ component of the 
Einstein equation reduces to 
\be
\label{eqn:aitcons2}
 \sum_i (\partial_{\bar{Z}} \tilde{a}_i)^2 = 0.
\ee
The $ZZ$ component of the Einstein equation gives
\be
\label{eqn:spCsol}
 C(Z,\bar{Z}) = \frac{1}{2} \log \partial_Z w + \nu + \tilde{c}(\bar{Z}),
\ee
where $\tilde{c}(\bar{Z})$ is arbitrary and
\be
\label{eqn:spdZnu}
 \partial_Z \nu = - \frac{w}{\partial_Z w} \sum_{i<j} \partial U_i
 \partial U_j.
\ee
The $Z \bar{Z}$ component of the Einstein equation is satisfied as a
consequence of these equations. The arbitrary function
$\tilde{c}(\bar{Z})$ reflects the freedom to do a coordinate
transformation $\bar{Z} \rightarrow \bar{Z}'(\bar{Z})$, and one can
also do a coordinate transformation $Z \rightarrow w(Z)$ to eliminate
$w(Z)$. Hence these solutions are characterized by $D-2$ functions $a_i(Z)$ 
and $D-2$ functions $\tilde{a}_i (\bar{Z})$. However, the constraints
\ref{eqn:aitcons1} and \ref{eqn:aitcons2} imply that only $D-4$ of the
functions $\tilde{a}_i$ are independent, and the constraint
\ref{eqn:aicons3} implies that only $D-3$ of the functions $a_i$ are
independent. For $D=4$, the solution is given by a single arbitrary
function $a_i(Z)$ and $\partial/\partial\bar{Z}$ is a null Killing
vector field so the solution describes a pp-wave spacetime
\cite{kramer:80}. These solutions admit a Killing spinor
\cite{tod:83}. The higher dimensional analogues of these pp-waves are
the solutions with $\tilde{a}_i = 0$ for all $i$, and it is
straightforward to show that these are the only Weyl solutions 
that satisfy the integrability conditions for the existence of a Killing 
spinor.

\sect{Weyl forms of flat space}
\label{app:flat}

The results of Appendix \ref{app:curvature} show that the Riemann
tensor of the metric \ref{eqn:lineelement} vanishes if, and only if,
\be
\label{eqn:flatcurv1}
 \partial_Z U_i \partial_{\bar{Z}} U_j + \partial_Z U_j
 \partial_{\bar{Z}} U_i = 0, \qquad i \ne j,
\ee
\be
\label{eqn:flatcurv2}
 \partial_Z^2 U_i + (\partial_Z U_i)^2 - 2 \partial_Z C \partial_Z U_i =
0,
\ee
\be
\label{eqn:flatcurv3}
 \partial_{\bar{Z}}^2 U_i + (\partial_{\bar{Z}} U_i )^2 - 2
 \partial_{\bar{Z}} C \partial_{\bar{Z}} U_i = 0,
\ee
\be
\label{eqn:flatcurv4}
 \partial_Z \partial_{\bar{Z}} U_i + \partial_Z U_i \partial_{\bar{Z}}
U_i = 0,
\ee
\be
\label{eqn:flatcurv5} 
 \partial_Z \partial_{\bar{Z}} C = 0.
\ee
Equation \ref{eqn:flatcurv4} can be immediately solved:
\be
\label{eqn:flatA}
 U_i(Z,\bar{Z}) = \log ( a_i(Z) + a_i(Z)^*),
\ee
where $a_i(Z)$ is arbitrary.

It is convenient to choose the labelling of the $U_i$ such
that $\partial_Z a_i \ne 0$ for $1 \le i \le r$ and $\partial_Z a_i = 0$
for $i>r$. Equation \ref{eqn:flatcurv2} is automatically satisfied if
$i>r$ or $j>r$. If $i,j \le r$ then this equation implies
\be
 \frac{\partial_Z a_i}{\partial_Z a_j} = -\frac{(\partial_Z a_i)^*}{
(\partial_Z a_j)^*} = i \lambda_{ij},
\ee
where $\lambda_{ij}$ is a real non-zero constant.
If $r>2$ then it follows that
\be
 \partial_Z a_3 = i\lambda_{32} \partial_Z a_2 = -\lambda_{32}
\lambda_{21} \partial_Z a_1.
\ee
However, this contradicts $\partial_Z a_3 = i \lambda_{31} \partial_Z
a_1$. Hence $r=0,1$ or $2$. These three cases will be discussed
individually. 

If $r=0$ then $a_i(Z)$ is a constant for all $i$ and hence $U_i$ is
constant for all $i$. Equations \ref{eqn:flatcurv2} and
\ref{eqn:flatcurv3} are trivially satisfied and equation
\ref{eqn:flatcurv5} has the solution
\be
 C(Z,\bar{Z}) = c(Z)+c(Z)^*,
\ee
where $c(Z)$ is arbitrary. This arbitrary function can be eliminated
by a coordinate transformation $Z \rightarrow Z'(Z)$. The line element
is then obviously flat. 

If $r = 1$ or $2$ then equations \ref{eqn:flatcurv2} and
\ref{eqn:flatcurv3} are trivially satisfied for $i>r$. For $i\le r$, the
solution is 
\be
 C(Z,\bar{Z}) = \frac{1}{2} \log \left( \partial_Z a_i \right) \left(
\partial_Z a_i \right)^* + {\rm constant},
\ee
which also ensures that equation \ref{eqn:flatcurv5} is satisfied. For
$r=1$, after changing coordinates from $Z$ to $a_1(Z)$, setting $a_1 =
\xi + i\eta$ and rescaling the coordinates to eliminate constants, 
the line element takes the form
\be
 ds^2 = \epsilon_1 \xi^2 (dx^1)^2 + \sum_{i=2}^{D-2}
 \epsilon_i (dx^i)^2 + d\xi^2 + d\eta^2.
\ee
This line element is
manifestly flat, with $x^1$ an angular coordinate (if $\epsilon_1 =
1$) or the boost coordinate in Rindler space (if $\epsilon_1 = -1$). 

For $r=2$, after changing and rescaling the coordinates as above, 
the line element takes the form 
\be
 ds^2 = \epsilon_1 \xi^2 (dx^1)^2 + \epsilon_2 \eta^2 (dx^2)^2 +
\sum_{i=3}^{D-2} \epsilon_i (dx^i)^2 + d\xi^2 + d\eta^2,
\ee
which is manifestly flat, with $x^1$ an angular/boost coordinate depending
on the sign of $\epsilon_1$, and similarly for $x^2$.

In order to identify the sources terms for Laplace's equation that
these flat metrics correspond to, one must first change coordinate from
$Z$ to $w(Z)$ (defined by equation \ref{eqn:wdef2} with $\tilde{w} =
w^*$). However, this is not possible if $r=0$ because then $w$ is
constant. Hence $r=0$ corresponds to the special case analyzed in
section \ref{sec:special}. 

For $r=1$, $U_i$ is constant for $i>1$ so equation \ref{eqn:wdef2} reduces to 
\be
 U_1 = \log (w+\bar{w}) + {\rm constant} = \log r + {\rm constant}.
\ee
where $w=r+iz$. The constant term can be eliminated by rescaling the
coordinate $x^1$. $U_1$ is the Newtonian potential produced by an 
infinite rod lying on the $z$-axis. The rod has vanishing thickness
and mass $1/2$ per unit length. 

For $r=2$, $U_i$ is constant for $i>2$. $U_1$ and $U_2$ are given by
equation \ref{eqn:flatA}. Recall that $\partial_Z a_2 = i \lambda_{21}
\partial_Z a_1$ hence $a_2 = \lambda_{21} (i a_1 + c)$, where $c$ is a
constant. The imaginary part of $c$ does not affect $U_2$ so $c$ can
be taken to be real. Equation \ref{eqn:wdef2} then gives
\be
 w+w^* \propto [a_1+a_1^*][i(a_1-a_1)^*+2c],
\ee
which can be solved to give $a_1$ in terms of $w$, and then express
$U_1$ and $U_2$ in terms of $w$:
\be
 U_1 = \log | \Re \sqrt{a \pm iw} | + {\rm constant},
\ee
\be 
 U_2 = \log | \Re \sqrt{-a \mp iw}| + {\rm constant},
\ee
where $a$ is an arbitrary real constant.

\sect{Formulae for the black ring}

\label{app:ring}

In order to write the metric of the black ring in Weyl form, it 
turns out to be convenient to look for constants $c$ and $\beta$ such that
\be
\label{eqn:square}
 r^2 + (z-c/A^2)^2 = \frac{1}{4(x-y)^2 A^4} \left( \alpha \mu x y - 2
cx -2 cy + \beta \right)^2.
\ee
{\it A priori}, one would not expect such constants to exist. However,
it turns out that they do, and are given by
\be
 c = \frac{\alpha^2 \mu}{\alpha \mu + \beta},
\ee
and
\be
 \beta^3 + \alpha \mu \beta^2 - \alpha^2 (\mu^2 + 4) \beta - \alpha^3
\mu (\mu^2 - 4) = 0.
\ee
The solutions are
\be
 \beta = \alpha \mu, \, c = \alpha/2, \qquad \beta = \alpha (2-\mu),
 \, c = \alpha \mu/2, \qquad \beta = -\alpha (2+\mu), \, c = -\alpha \mu/2.
\ee
Note that the solutions for $c$ can be written as $A^2 a_i$, where the
quantities $a_i$ were defined in section \ref{sec:ring}. It follows
that each function $R_i$ defined in section \ref{sec:ring} is 
the square root of the right hand side of equation \ref{eqn:square}
for the appropriate value of $c$. Since the right hand side of this
equation is a perfect square, the expressions for $R_i$ turn out to be
quite simple:
\be
 R_1 = \frac{\alpha (\mu x y - x - y + \mu)}{2 A^2 (x-y)},
\ee
\be
 R_2 = \frac{\alpha (\mu x y - \mu x - \mu y + 2 - \mu) }{2 A^2
(x-y)},
\ee
\be
 R_3 = \frac{\alpha (-\mu x y - \mu x - \mu y + 2 + \mu)}{2 A^2
(x-y)}.
\ee
It is also possible to show
\be
 R_1 + \zeta_1 = \frac{\alpha (1-x^2) F(y)}{A^2 (x-y)^2},
\ee
\be
 R_1 - \zeta_1 = \frac{\alpha (y^2-1) F(x)}{A^2 (x-y)^2},
\ee
\be
 R_2 + \zeta_2 = \frac{\alpha (1+x) (1-y) F(x)}{A^2(x-y)^2},
\ee
\be
 R_2 - \zeta_2 = \frac{\alpha (1-x) (-1-y) F(y)}{A^2(x-y)^2},
\ee
\be
 R_3 + \zeta_3 = \frac{\alpha (1+x)(1-y) F(y)}{A^2(x-y)^2},
\ee
\be
 R_3 - \zeta_3 = \frac{\alpha (1-x)(-1-y) F(x)}{A^2(x-y)^2},
\ee
\be
 Y_{12} = \frac{\alpha^2 (1-x)(1-y)F(x)F(y)}{2 A^4 (x-y)^2},
\ee
\be
 Y_{13} = \frac{\alpha^2 (1+\mu)^2 (1-x)(1-y)}{2 A^4 (x-y)^2},
\ee
\be
 Y_{23} = \frac{2\alpha^2  F(x)F(y)}{A^4 (x-y)^2}.
\ee 

\sect{Calculating $\nu$ integrals}

\label{app:integrate}

The purpose of this appendix is to explain how the quantity $\nu$ in
equation \ref{eqn:weylmetric} is calculated for the solutions of
section \ref{sec:newsol}. Consider first the solution of section
\ref{sec:holebubble}. In terms of the complex coordinates $w$, the
functions $U_i$ take the form
\be
 U_1 = 
 \log |\Re \left[ (a_2 + iw)^{1/2} \right]| - \log |\Re \left[ (a_1
+ iw)^{1/2} \right]|,
\ee
\be
 U_2 = \log |\Re \left[ (a_3 + iw)^{1/2} \right]| - \log |\Re \left[
(a_2 + iw)^{1/2} \right]|,
\ee
\be
 U_3 = \log |\Re \left[ (a_1 + iw)^{1/2} \right]| + \log |\Im \left[
(a_3
+ iw)^{1/2} \right]|,
\ee
where the constants $a_i$ are defined in equation \ref{eqn:aidef} and
arbitrary additive constants are suppressed. To calculate $\nu$
directly, these expressions could be substituted into equations
\ref{eqn:dwnu} and \ref{eqn:dwbarnu}. On the right hand side of these
equations there would be $12$ terms ($4$ from each $\partial U_i
\partial U_j$).
Alternatively, one can instead deal with $\gamma$, which is determined 
by equations \ref{eqn:dwgamma} and \ref{eqn:dwbargamma}, with only 10 
terms on the right hand side. Having obtained $\gamma$, $\nu$ can be
immediately calculated using
\be
 \nu = \gamma - \sum_i U_i.
\ee
Equation \ref{eqn:dwgamma} takes the form
\ba
\label{eqn:dgamma}
 \partial \gamma &=& \frac{1}{4w} + F_+(w,\bar{w};a_1) +
 F_+(w,\bar{w};a_2) + F_+(w,\bar{w};a_3) + F_-(w,\bar{w};a_3)
\nonumber \\ 
 &-& 2G_{++}(w,\bar{w};a_2,a_3) + 2 G_{-+}(w,\bar{w};a_1,a_3) \\ &+&
 G_{++}(w,\bar{w};a_1,a_3) - G_{++}(w,\bar{w};a_1,a_2) -
 G_{-+}(w,\bar{w};a_2,a_3), \nonumber
\ea
where the functions $F$ and $G$ are defined by
\be
 F_{\pm}(w,\bar{w};c) = -\frac{w+\bar{w}}{4 (c+iw) \left[(c+iw)^{1/2}
 \pm (c-i\bar{w})^{1/2} \right]^2},
\ee
\ba
 &{}&G_{\pm \pm'}(w,\bar{w};c,d) = \\ {}&-& \frac{w+\bar{w}}{4 (c+iw)^{1/2}
(d+iw)^{1/2} \left[(c+iw)^{1/2} \pm' (c-i\bar{w})^{1/2} \right] \left[
(d+iw)^{1/2} \pm (d-i\bar{w})^{1/2} \right]}, \nonumber
\ea
where $c$ and $d$ are real constants with $c > d$. In order to
integrate equation \ref{eqn:dgamma}, it is necessary to integrate $F$
and $G$. This can be done by a change of variable to $Z =
(c+iw)^{1/2}$, which yields
\be
 \int dw F_{\pm} (w,\bar{w};c) = -\frac{1}{4} \log (c+iw) + \log \left[
 (c+iw)^{1/2} \pm (c-i\bar{w})^{1/2} \right] + \ldots,
\ee
where the ellipsis denotes an arbitrary function of $\bar{w}$. One
similarly obtains
\ba
 \int dw G_{++}(w,\bar{w};c,d) &=& \log \Re \left[ (c+iw)^{1/2}
  + (d+iw)^{1/2} \right]  \nonumber \\ &-& 
 \frac{1}{2} \log \left[ (c+iw)^{1/2} +
 (d+iw)^{1/2} \right] + \ldots 
\ea
\ba
 \int dw G_{--} (w,\bar{w};c,d) &=& \log \Im \left[ (c+iw)^{1/2} +
 (d+iw)^{1/2} \right]  \nonumber \\ &-& 
 \frac{1}{2} \log \left[ (c+iw)^{1/2} +
 (d+iw)^{1/2} \right] + \ldots,
\ea
\ba
 \int dw G_{+-} (w,\bar{w};c,d) &=& \log \Re \left[ (d+iw)^{1/2}
 \right]  - \log \Re \left[ (c+iw)^{1/2} + (d+iw)^{1/2} \right]
 \nonumber \\ &+& \frac{1}{2} \log \left[ (c+iw)^{1/2} + (d+iw)^{1/2}
 \right] + \ldots,
\ea
\ba
 \int dw G_{-+} (w,\bar{w};c,d) &=& \log \Im \left[ (d+iw)^{1/2}
 \right] -\log \Im \left[ (c+iw)^{1/2} + (d+iw)^{1/2} \right]
 \nonumber \\ &+& \frac{1}{2} \log \left[ (c+iw)^{1/2} + (d+iw)^{1/2}
 \right] + \ldots.
\ea
These results yield $\gamma$, and hence $\nu$, up to an arbitrary
function of $\bar{w}$. This function can be determined up to a real
constant of integration by demanding that $\nu$ be real. Finally, the
following expressions can be used to express $\nu$ in terms of $R_i$,
$\zeta_i$ and $Y_{ij}$:
\be
 R_i = |a_i + iw|,
\ee
\be
 R_i - \zeta_i = |\Re \left[(a_i + iw)^{1/2} \right]|,
\ee
\be
 R_i + \zeta_i = |\Im \left[(a_i + iw)^{1/2} \right]|,
\ee
\ba
 Y_{ij} &=& 2 \log |\Re \left[ (a_i +iw)^{1/2} \right] | + 2\log | \Re
 \left[ (a_j + iw)^{1/2} \right] | \\ &+&  4 \log |(a_i +
 iw)^{1/2} + (a_j + iw)^{1/2} | - 4 \log | \Re \left[ (a_i + iw)^{1/2} + (a_j
 + iw)^{1/2} \right]|, \nonumber
\ea 
where additive constants have again been suppressed. The following
identity is useful in rearranging $\nu$ so that it can be written in
terms of the above expressions:
\ba
 \log |\Im \left[(c+iw)^{1/2} + (d+iw)^{1/2} \right]| - \log |\Im \left[
 (d+iw)^{1/2} \right]| \nonumber \\
 = \log |\Re \left[(c+iw)^{1/2} + (d+iw)^{1/2} \right]|
 - \log |\Re \left[ (c+iw)^{1/2} \right]|.
\ea

\end{document}